\documentclass[preprint,aps]{revtex4-1}
\pdfoutput=1

\usepackage{amsmath,amssymb,amsfonts,dcolumn,color,graphicx,graphics,latexsym,placeins,epsfig}
\usepackage{epsfig}
\usepackage{bm}
\usepackage{slashed}
\usepackage{latexsym}
\usepackage{natbib}
\usepackage{url}
\usepackage{dcolumn}
\usepackage{color}
\usepackage{amsfonts,amssymb,amsmath}
\usepackage{graphicx,epsfig}
\usepackage{psfrag}
\usepackage{subfigure}
\usepackage{tabularx}
\usepackage{hyperref}
\hypersetup{colorlinks=true}

\newcommand{\be}{\begin{equation}}
\newcommand{\ee}{\end{equation}}
\newcommand{\ba}{\begin{eqnarray}}
\newcommand{\ea}{\end{eqnarray}}

\DeclareUnicodeCharacter{2212}{\textendash}

\begin{document}

\title{Energizing gamma ray bursts via $Z^\prime$ mediated neutrino heating }
 \author{Tanmay Kumar Poddar}
\email[Email Address: ]{tanmay.poddar@tifr.res.in}
\affiliation{Theoretical Physics Division, 
Physical Research Laboratory, Ahmedabad - 380009, India}
\affiliation{Department of Theoretical Physics, Tata Institute of Fundamental Research, Mumbai - 400005, India}

\author{Srubabati Goswami}
\email[Email Address: ]{sruba@prl.res.in}
\affiliation{Theoretical Physics Division, 
Physical Research Laboratory, Ahmedabad - 380009, India}

\author{Arvind Kumar Mishra }
\email[Email Address: ]{arvind.mishra@acads.iiserpune.ac.in}
\affiliation{Indian Institute of Science Education and Research, Pune 411008, India}
\affiliation{ Theoretical Physics Division, 
Physical Research Laboratory, Ahmedabad - 380009, India}


\begin{abstract}
The pair annihilation of neutrinos $(\nu\overline{\nu}\rightarrow e^+e^-)$ can energize violent stellar explosions such as gamma ray bursts (GRBs). The energy in this neutrino heating mechanism can be further enhanced by modifying the background spacetime over that of Newtonian spacetime. However, one cannot attain the maximum GRB energy $(\sim 10^{52}~\rm{erg})$ in either the Newtonian background or Schwarzschild and Hartle-Thorne background. On the other hand, using modified gravity theories or the Quintessence field as background geometries, the maximum GRB energy can be reached. In this paper, we consider extending the standard model by an extra $U(1)_{\rm{B-L}}$ gauge group and augmenting the energy deposition by neutrino pair annihilation process including contributions mediated by the $Z^\prime$ gauge boson belonging to this model. From the observed energy of GRB, we obtain constraints on $U(1)_{\rm{B-L}}$ gauge coupling in different background spacetimes. We find that the bounds on gauge coupling in modified gravity theories and quintessence background are stronger than those coming from the neutrino-electron scattering experiments in the limit of small gauge boson masses. Future GRB observations with better accuracy can further strengthen these bounds.

\end{abstract}

\maketitle
\section{Introduction}
It has long been realized that the neutrino pair annihilation process plays a significant role in depositing energy to violent stellar processes such as type II supernovae \cite{ Bethe:1990mw,Goodman:1986we,Bethe:1985sox}, merging neutron stars \cite{Bhattacharya1991}, binary neutron stars in the last stable orbit \cite{Mathews:1997vw}, Gamma Ray Bursts (GRBs) \cite{Salmonson:1999es,Salmonson:2001tz,Asano:2000ib,Asano:2000dq,Prasanna:2001ie,Birkl:2006mu,Miller:2002hqu,Meszaros:1992gc,Ruffert:1998qg,Zhang:2009ew,Kovacs:2010zp,Lambiase:2020iul,Lambiase:2020pkc,Paczynski1990,Cooperstein1987}, etc.
It is believed that most of the energy in such stellar explosions is carried away by the three flavors of neutrinos. The emission of a huge number of neutrinos with luminosity $L_\nu\sim 10^{53}~\rm{erg/s}$ makes the stellar objects cool. 
A fraction of such huge neutrino flux can also deposit energy into the stellar envelope through neutrino pair annihilation $(\nu_i\overline{\nu}_i\rightarrow e^+e^-, i=e, \mu, \tau)$, neutrino lepton scattering, and neutrino baryon capture \cite{Lambiase:2020iul}. This mechanism is termed neutrino heating. The process $\nu_i\overline{\nu}_i\rightarrow e^+e^-$ is important for collapsing neutron stars, binary neutron stars in their last stable orbit, and r-process nucleosynthesis \cite{Eichler:1989ve}. This annihilation process can also continuously give energy to the radiation bubble and is a possible source of powering GRBs \cite{Salmonson:1999es}. In fireball model \cite{Leng:2014dfa}, the subsequent annihilation of electron-positron pair into photons ultimately power the GRB. Here, we assume that all the energy released into electrons is converted into photons to get such high energy in GRBs \cite{Salmonson:1999es}. These GRBs are high energy explosions that have been observed from cosmological distances \cite{Zhang:2003uk,Piran:2004ba,Luongo:2021pjs,Meszaros:1992gc}. They are the most energetic phenomenon represented by intense and prompt $\gamma$-ray emissions. We consider the maximum energy emission associated with the short GRBs to be $\sim 10^{52}~\rm{erg}$ \cite{Perego:2017fho}. However, in a Newtonian background, the neutrino pair annihilation process cannot provide the maximum energy.

The effect of including different background spacetimes near such strong gravity regimes has also been underscored in the literature. This is also apparent from the values of  $\frac{2GM}{R}$ which for instance is $\sim 0.7$ for collapsing neutron stars, and $\sim 0.4$ for supernovae calculations and hence, one cannot neglect the effect of General Relativity (GR) in the strong gravity regime \cite{Mathews:1997vw}. Here, $G$ denotes Newton's gravitational constant, $M$ denotes the mass of the neutron star, and $R$ denotes the length scale. Since the neutron stars are rotating, the rotation parameter in the background metric also has to be included. 
It has been shown that in the Schwarzschild background, the neutrino heating is enhanced up to a factor $4$ for type II supernova, and by up to a factor $30$ for collapsing neutron stars relative to the Newtonian result \cite{Salmonson:1999es}. The rotation on the other hand reduces the energy deposition by $38\%$ compared to the non rotating case \cite{Prasanna:2001ie}. However, in all three backgrounds (Newtonian, Schwarzschild, and Hartle-Thorne) one cannot attain the maximum GRB energy. The energy deposition due to neutrino heating can also be calculated in different modified gravity backgrounds such as Born-Infeld Reissner-Nordstrom gravity \cite{Breton:2002td}, charged Galileon gravity \cite{Babichev:2015rva}, Eddington inspired Born-Infeld gravity \cite{BeltranJimenez:2017doy}, Einstein dilaton Gauss-Bonnet gravity \cite{Sotiriou:2014pfa}, Brans-Dicke gravity \cite{Brans:1961sx}, higher derivative gravity \cite{Kokkotas:2017zwt}, etc as it is done in \cite{Lambiase:2020iul}. In some of the modified gravity theories such as the Reissner-Nordstrom solution in the Born-Infeld model \cite{Breton:2002td}, one can attain the maximum GRB energy. The quintessence background can also enhance the energy deposition to its maximum value as discussed in \cite{Lambiase:2020pkc}. The radial variation of the temperature for black hole accretion disk in modified gravity theories can enhance the energy deposition by one order magnitude with respect to GR as is recently discussed in \cite{Lambiase:2022ywp}. The energy deposition can be enhanced as well due to the presence of topological defects such as global monopole \cite{Shi:2022pbc}. All these studies have considered only the Standard Model (SM) contribution to the neutrino pair annihilation process. 
  

In this paper, we consider for the first time, the possibility of additional contribution from BSM physics to the neutrino pair annihilation process in the context of explaining the observed energy budget of GRBs. In particular, we focus on the scenario where the SM is extended by a general $U(1)_X$ gauge group. Such a scenario contains an additional neutral gauge boson ($Z^\prime$) associated with the $U(1)_X$ symmetry. The special case of $U(1)_X$ is the $U(1)_{B-L}$ model which has been extensively studied in the literature \cite{Das:2021nqj,Chakraborty:2021apc,Feng:2022inv}. The phenomenology of extra $Z^\prime$ in these models is particularly interesting and constraints have been obtained on the mass and the coupling strength of $Z^\prime$ from several experiments. This includes electroweak precession data \cite{Erler:2009jh} , collider searches \cite{Dittmar:2003ir,Basso:2008iv,Das:2016zue,Accomando:2017qcs,Ekstedt:2016wyi,Das:2019pua,ATLAS:2017fih,ATLAS:2017rue} , neutrino-electron scattering experiments \cite{PhysRevD.92.033009,Lindner2018}, beam dump experiments \cite{Alekhin:2015byh,PhysRevD.99.095011,FASER:2019aik}, SN1987A \cite{Dent:2012mx,Raffelt:2000kp,Kazanas:2014mca,Balaji:2022noj} etc. If the $Z^\prime$ gauge boson has a coupling with neutrinos and electrons then the $Z^\prime$ mediated neutrino pair annihilation process can also contribute to the neutrino pair annihilation process. From the reported energy of GRB $(10^{52}~\rm{erg})$ \cite{Perego:2017fho,Fong:2015oha}, we obtain constraints on gauge coupling in Newtonian, Schwarzschild, Hartle-Thorne, Born-Infeld Reissner-Nordstrom gravity, and Quintessence backgrounds.

The paper is organized as follows. In Section \ref{sec2}, we obtain the total cross section and the energy deposition rate due to $Z^\prime$ mediated process in neutrino pair annihilation. In Section \ref{sec3} we compute the angular integration in Newtonian, Schwarzschild, Hartle-Thorne, Born-Infeld Reissner-Nordstrom gravity, and Quintessence spacetime backgrounds for the neutrino heating process. The contribution of $Z^\prime$ mediated process to the neutrino heating mechanism in different spacetime backgrounds is calculated in Section \ref{sec4}. In Section \ref{sec5}, we obtain the amount of energy enhancement due to $Z^\prime$ contribution via neutrino heating process for the above mentioned spacetime backgrounds. In Section \ref{combined}, we obtain constraints on $Z^\prime$ from the GRB observation. Finally, in Section \ref{sec6}, we conclude and discuss our results.

In the following, we have used natural units $(c=1, \hbar=1)$, and $G=1$ throughout the paper.
\section{Neutrino heating through $Z^\prime$}\label{sec2}
We extend the SM gauge group $(SU(3)_c\times SU(2)_L\times U(1)_Y)$ by an additional $U(1)_X$ gauge symmetry. Such a scenario includes an extra neutral gauge boson ($Z^\prime$) associated with the $U(1)_X$ symmetry. The latter is broken by an extra singlet scalar field $\Phi$ and the gauge boson acquires mass. Three right handed neutrinos are needed in the model to cancel the gauge and gauge-gravity anomalies. The $U(1)_X$ charges of the quarks and the leptons can be expressed in terms of that of the $\Phi$ and the SM Higgs ($H$). The charges for the scalars are chosen as $2 x_\Phi$ and $\frac{x_H}{2}$ \cite{Das:2021nqj}. The corresponding $U(1)_X$ charge of lepton doublet is $Q^l_X=(-\frac{1}{2}x_H-x_\Phi)$ and the charges of right handed electron and neutrino are $Q^{e_R}_X=(-x_H-x_\Phi)$ and $Q^{N_R}_X=-x_\Phi$ respectively. The particular choice of $x_H=0$ and $x_\Phi=1$ leads to the $U(1)_{\rm{B-L}}$ model \cite{PhysRevD.20.776,Marshak:1979fm,Mohapatra:1980qe,Wetterich:1981bx,Masiero:1982fi}. The general interaction Lagrangian of $Z^\prime$ gauge bosons with the leptons is 
\begin{equation}
-\mathcal{L}_{int}\supset g^\prime(Q^l_X\overline{l_L}\gamma^\mu l_LZ^\prime_\mu+Q^{e_R}_X\overline{l_R}\gamma^\mu l_RZ^\prime_\mu).
\label{eq:1}
\end{equation}
The electron neutrino contributes to the neutrino annihilation process $\nu_e\overline{\nu}_e\rightarrow e^+e^-$ via charge current $(W)$, neutral current $(Z)$, and $Z^\prime$ mediated interactions whereas $\nu_\mu$ and $\nu_\tau$ have only $Z$ and $Z^\prime$ mediated interactions for the process $\nu_{\mu,\tau}\overline{\nu}_{\mu,\tau}\rightarrow e^+e^-$.
\begin{figure}[h]
\includegraphics[height=6cm]{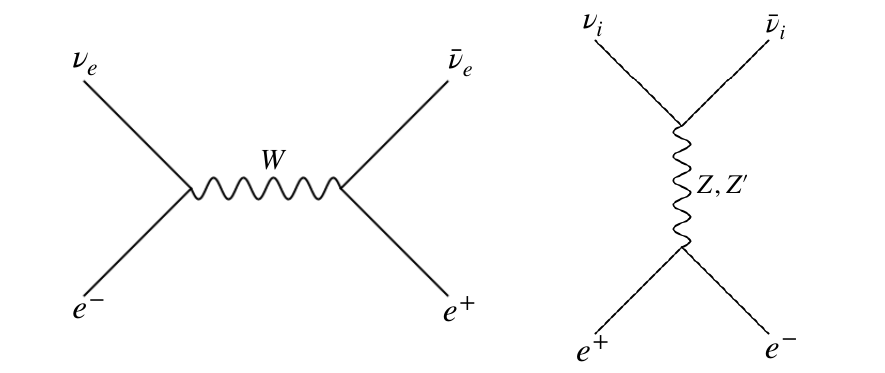}
\caption{Feynman diagrams contributing to the process $\nu_i\overline{\nu}_i\rightarrow e^+e^-$.}
\label{feynman}
\end{figure} 
Since all three flavors of neutrinos are there in a hot neutron star, they will all contribute to the energy deposition rate via the Feynman diagrams Fig.\ref{feynman}. The energy deposition rate per unit volume near a hot neutron star is given as \cite{Goodman:1986we}
\begin{equation}
\dot{q}(r)=\int \int f_\nu(\textbf{p}_\nu, r) f_{\overline{\nu}}(\textbf{p}_{\overline{\nu}}, r)(\sigma |\textbf{v}_\nu -\textbf{v}_{\overline{\nu}}|E_\nu E_{\overline{\nu}})\times \frac{E_\nu+E_{\overline{\nu}}}{E_\nu E_{\overline{\nu}}}d^3\textbf{p}_\nu d^3\textbf{p}_{\overline{\nu}},
\label{eq:2}
\end{equation} 
where $E_\nu$ denotes the energy of neutrino and $f_{\nu}=\frac{2}{(2\pi)^3}(e^{E_{\nu}/kT}+1)^{-1}$ corresponds to their thermal energy distribution function in the phase space which is of Fermi-Dirac type. Here $k$ denotes the Boltzmann constant and $T$ denotes the neutrino temperature. The neutrino velocity is denoted as $v_\nu$ and $\sigma$ denotes the cross section in the rest frame. Eq.\ref{eq:2} is true for any flavor of neutrinos. Since the term in the first bracket of Eq.\ref{eq:2} is a Lorentz invariant quantity, we can calculate its value in the centre of mass frame for $\nu_e\overline{\nu}_e\rightarrow e^+e^-$ process. In the $U(1)_X$ model, it can be expressed as 
\begin{equation}
\begin{split}
(\sigma |\textbf{v}_{\nu_e} -\textbf{v}_{{\overline{\nu}}_e}|E_{\nu_e} E_{{\overline{\nu}}_e})_{U(1)_X}=\Big[\frac{G^2_F}{3\pi}(1+4\sin^2\theta_W+8\sin^4\theta_W)+\frac{4g^{\prime^4}}{6\pi M_{Z^\prime}^4}\Big\{\Big(\frac{3}{4}x_H+x_\Phi\Big)^2+\Big(\frac{x_H}{4}\Big)^2\Big\}\times\\
\Big\{\Big(x_\Phi+\frac{x_H}{4}\Big)^2+\Big(\frac{x_H}{4}\Big)^2\Big\}+
\frac{4G_Fg^{\prime^2}}{3\sqrt{2}\pi M^2_{Z^\prime}}\Big(x_\Phi+\frac{x_H}{2}\Big)\Big[\Big(\frac{3}{4}x_H+x_\Phi\Big)\Big(-\frac{1}{2}+2\sin^2\theta_W\Big)+\frac{x_H}{8}\Big]+\\
\frac{4G_Fg^{\prime^2}}{3\sqrt{2}\pi M_{Z^\prime}^2 }\Big(x_\Phi+\frac{x_H}{2}\Big)^2\Big](E_{\nu_e} E_{\overline{\nu}_e}-\textbf{p}_{\nu_e}.\textbf{p}_{\overline{\nu}_e})^2,
\end{split}
\label{eq:3}
\end{equation}
where for the region of interest we neglect the mass of the electron, since, the energy of neutrino is greater than $10\hspace{0.1cm}\rm{MeV}$; $G_F=1.166\times 10^{-5}\hspace{0.1cm}\rm{GeV^{-2}}$ is the Fermi constant and the $\theta_W$ is the Weinberg angle whose value is $\sin^2\theta_W=0.23$ \cite{ParticleDataGroup:2020ssz}.
The first term in Eq.\ref{eq:3} corresponds to the $W$ and $Z$ mediated SM processes, the second term is due to the $Z^\prime$ contribution only. The third term arises because of the interference between $Z$ and $Z^\prime$ mediated diagrams, while the fourth term stems from the interference between the $W$ and $Z^\prime$ mediated processes. For muon and tau types of neutrinos, only $Z$ and $Z^\prime$ mediated processes will contribute. Hence, for the scattering $\nu_{\mu,\tau}\overline{\nu}_{\mu,\tau}\rightarrow e^+e^-$, we obtain
\begin{equation}
\begin{split}
(\sigma |\textbf{v}_{\nu_{\mu,\tau}} -\textbf{v}_{{\overline{\nu}}_{\mu,\tau}}|E_{\nu_{\mu,\tau}} E_{{\overline{\nu}}_{\mu,\tau}})_{U(1)_X}=\Big[\frac{G^2_F}{3\pi}(1-4\sin^2\theta_W+8\sin^4\theta_W)+\frac{4g^{\prime^4}}{6\pi M_{Z^\prime}^4}\Big\{\Big(\frac{3}{4}x_H+x_\Phi\Big)^2+\Big(\frac{x_H}{4}\Big)^2\Big\}\times\\
\Big\{\Big(x_\Phi+\frac{x_H}{4}\Big)^2+\Big(\frac{x_H}{4}\Big)^2\Big\}+
\frac{4G_Fg^{\prime^2}}{3\sqrt{2}\pi M^2_{Z^\prime}}\Big(x_\Phi+\frac{x_H}{2}\Big)\Big[\Big(\frac{3}{4}x_H+x_\Phi\Big)\Big(-\frac{1}{2}+2\sin^2\theta_W\Big)+\frac{x_H}{8}\Big]\Big]\times\\
(E_{\nu_{\mu,\tau}} E_{\overline{\nu}_{\mu,\tau}}-\textbf{p}_{\nu_{\mu,\tau}}.\textbf{p}_{\overline{\nu}_{\mu,\tau}})^2.
\end{split}
\label{eq:5}
\end{equation}
In what follows, we focus on the $U(1)_{B-L}$ model which has fixed values of $x_H$ and $x_\Phi$ as $x_H=0$ and $x_\Phi=1$. The results can easily be generalized to the $U(1)_X$ case by using suitable values of $x_H$ and $x_\Phi$.

Putting $P_\nu=E_\nu \Omega_\nu$, and $d^3\textbf{p}_\nu=E^2_\nu dE_\nu d\Omega_\nu$ in the direction of $\Omega_\nu$, where $d\Omega_\nu$ denotes the solid angle, and assuming $T_\nu=T_{\overline{\nu}}=T$ we find
\begin{equation}
\int\int f_\nu f_{\overline{\nu}}(E_\nu+E_{\overline{\nu}})E^3_\nu E^3_{\bar{\nu}}dE_\nu dE_{\overline{\nu}}=\frac{21}{2(2\pi)^6}\pi^4(kT)^9\zeta(5).
\label{eq:7}
\end{equation}
 Hence, from Eq.\ref{eq:2} the energy deposition rate due to the electron neutrino pair annihilation process in $U(1)_{\rm{B-L}}$ model becomes
\begin{equation}
\begin{split}
\dot{q}_{\nu_e}(r)=\frac{21}{2(2\pi)^6}\pi^4(kT_{\nu_e}(r))^9\zeta(5)\times\Big[\frac{G^2_F}{3\pi}(1+4\sin^2\theta_W+8\sin^4\theta_W)+\frac{4g^{\prime^4}}{6\pi M_{Z^\prime}^4}+\\
\frac{4 G_F g^{\prime^2}}{3\sqrt{2}\pi M^2_{Z^\prime}}\Big(-\frac{1}{2}+2\sin^2\theta_W\Big)+\frac{4G_Fg^{\prime^2}}{3\sqrt{2}\pi M_{Z^\prime}^2 }\Big]\Theta_{\nu_e}(r),
\end{split}
\label{eq:8}
\end{equation}
where the angular integration $\Theta(r)$ is defined as 
\begin{equation}
\Theta(r)=\int\int (1-\Omega_\nu.\Omega_{\overline{\nu}})^2 d\Omega_\nu d\Omega_{\overline{\nu}}.
\label{eq:9}
\end{equation}
Similarly the energy deposition due to muon and tau type neutrino annihilations in $U(1)_{\rm{B-L}}$ model is given by
\begin{equation}
\begin{split}
\dot{q}_{\nu_{\mu,\tau}}(r)=\frac{21}{2(2\pi)^6}\pi^4(kT_{\nu_{\mu,\tau}}(r))^9\zeta(5)\times\Big[\frac{G^2_F}{3\pi}(1-4\sin^2\theta_W+8\sin^4\theta_W)+\frac{4g^{\prime^4}}{6\pi M_{Z^\prime}^4}+\\
\frac{4 G_F g^{\prime^2}}{3\sqrt{2}\pi M^2_{Z^\prime}}\Big(-\frac{1}{2}+2\sin^2\theta_W\Big)\Big]\Theta_{\nu_{\mu,\tau}}(r).
\end{split}
\label{eq:10}
\end{equation}
In the SM limit $(\frac{g^\prime}{M_{Z^\prime}}\rightarrow 0)$, we get back the earlier result \cite{Cooperstein1986,Cooperstein1987}
\begin{equation}
\dot{q}(r)=\frac{7G^2_F\pi^3\zeta(5)}{2(2\pi)^6}(kT)^9\Theta(r)(1\pm 4\sin^2\theta_W+8\sin^4\theta_W),
\label{eq:11}
\end{equation}
where the $+$ sign is for $\nu_e\overline{\nu}_e$ pair and the $-$ sign is for $\nu_\mu\overline{\nu}_\mu$ and $\nu_\tau\overline{\nu}_\tau$ pairs.
\section{Neutrino heating in Different background spacetimes }\label{sec3}
In this section, we calculate the effect of geometries in neutrino heating. We mainly focus on Hartle-Thorne (HT) geometry, the Born-Infeld generalization of Reissner-Nordstrom geometry (BIRN), and Quintessence geometry (Quint). We also calculate the neutrino heating in different limiting cases such as Newtonian (Newt) and Schwarzschild (Sch) backgrounds.
\subsection{Hartle-Thorne background}
The geodesic outside a slowly rotating neutron star with only a dipole correction on a static star is governed by the HT metric \cite{Hartle:1968si}
\begin{equation}
ds^2=-\Big(1-\frac{2M}{r}+\frac{2J^2}{r^4}\Big)dt^2+\Big(1-\frac{2M}{r}+\frac{2J^2}{r^4}\Big)^{-1}dr^2+r^2d\theta^2+\Big(d\phi-\frac{2J}{r^3}dt\Big)^2,
\label{eq:12}
\end{equation}
where $r$ denotes the distance from the origin, $\phi$ is the longitude, $M$ is the mass of the neutron star, and $J$ is the specific angular momentum. For $J=0$, Eq.\ref{eq:12} reduces to the Schwarzschild metric and for both $J=M=0$, we have the flat or the Newtonian metric. Since, we have considered the planar motion for the massless particle so we take $\theta=\frac{\pi}{2}$ and the null geodesic $g_{\mu\nu}V^\mu V^\nu=0$ , where $V^\mu=\frac{dX^\mu}{d\lambda}$ and $X^\mu\equiv(t, r, \theta, \phi)$. Hence, the geodesic equation of the HT metric for the planar motion becomes 
\begin{equation}
-\Big(1-\frac{2M}{r}-\frac{2J^2}{r^4}\Big)\dot{t}^2+\Big(1-\frac{2M}{r}+\frac{2J^2}{r^4}\Big)^{-1}\dot{r}^2+r^2\dot{\phi}^2-\frac{4J}{r}\dot{\phi}\dot{t}=0.
\label{eq:13}
\end{equation}
We can also derive the generalized momenta as
\begin{equation}
p_t=-\Big(1-\frac{2M}{r}-\frac{2J^2}{r^4}\Big)\dot{t}-\frac{2J}{r}\dot{\phi}=-E, \hspace{0.2cm} p_r=\Big(1-\frac{2M}{r}+\frac{2J^2}{r^4}\Big)^{-1}\dot{r}, \hspace{0.2cm} p_\phi=-\frac{2J\dot{t}}{r}+r^2\dot{\phi}=L,
\label{eq:14}
\end{equation}
where $E$ denotes the energy per unit mass of the system and $L$ denotes the angular momentum per unit mass. We can solve Eq.\ref{eq:14} and obtain the expressions of $\dot{t}$ and $\dot{\phi}$ in terms of $L$ and $E$ as
\begin{equation}
\dot{t}=\Big(E-\frac{2JL}{r^3}\Big)\Big(1-\frac{2M}{r}+\frac{2J^2}{r^4}\Big)^{-1}, \hspace{0.2cm} \dot{\phi}=\Big[\frac{L}{r^2}\Big(1-\frac{2M}{r}-\frac{2J^2}{r^4}\Big)+\frac{2JE}{r^3}\Big]\Big(1-\frac{2M}{r}+\frac{2J^2}{r^4}\Big)^{-1}.
\label{eq:15}
\end{equation}
Using Eq.\ref{eq:14}, we can write Eq.\ref{eq:13} as 
\begin{equation}
-E\dot{t}+L\dot{\phi}+\Big(1-\frac{2M}{r}+\frac{2J^2}{r^4}\Big)^{-1}\dot{r}^2=0.
\label{eq:16}
\end{equation}
Putting the expressions of $\dot{t}$ and $\dot{\phi}$ from Eq.\ref{eq:15}, we can write Eq.\ref{eq:16} as
\begin{equation}
\frac{1}{r^4}\Big(\frac{dr}{d\phi}\Big)^2=\frac{\Big(1-\frac{2M}{r}+\frac{2J^2}{r^4}\Big)^2}{\Big(1-\frac{2M}{r}-\frac{2J^2}{r^4}+\frac{2JE}{Lr}\Big)^2}\Big[\frac{E^2}{L^2}-\frac{4JE}{Lr^3}-\frac{1}{r^2}\Big(1-\frac{2M}{r}-\frac{2J^2}{r^4}\Big)\Big].
\label{eq:17}
\end{equation}
The local Lorentz tetrad for the given metric can be written as \cite{Prasanna:2001ie}
\begin{equation}
e^\mu_i=
\begin{bmatrix}
\Big(1-\frac{2M}{r}+\frac{2J^2}{r^4}\Big)^\frac{1}{2} & 0 & 0 &0\\
0 & \Big(1-\frac{2M}{r}+\frac{2J^2}{r^4}\Big)^{-\frac{1}{2}} &0 &0\\
0&0&r&0\\
-\frac{2J}{r^2}&0&0& r\sin\theta
\end{bmatrix},
\label{eq:18}
\end{equation} 
where the upper index denotes the row and the lower index denotes the column. The tangent of the angle $(\theta_r)$ between the trajectory and the tangent vector is the ratio between the radial and longitudinal velocity that can be expressed in terms of $\frac{dr}{d\phi}$ as 
\begin{equation}
\Big(\frac{dr}{d\phi}\Big)=\Big(1-\frac{2M}{r}+\frac{2J^2}{r^4}\Big)^\frac{3}{2}\Big[\Big(1-\frac{2M}{r}-\frac{2J^2}{r^4}\Big)\frac{1}{r}+\frac{2JE}{Lr^2}\Big]^{-1}\tan\theta_r.
\label{eq:19}
\end{equation}
Comparing Eq.\ref{eq:17} and Eq.\ref{eq:19} we obtain the impact parameter $b_I=\frac{L}{E}$ as
\begin{equation}
b_I=\Bigg[\frac{2J}{r^3}+\frac{\Big(1-\frac{2M}{r}+\frac{2J^2}{r^4}\Big)^\frac{1}{2}}{r\cos\theta_r}\Bigg]^{-1}.
\label{eq:20}
\end{equation}
If the neutrino is emitted tangentially $(\theta_R=0)$ from the neutrinosphere of radius $R_{\nu_i}$, then its trajectory defined by an angle $\theta_r$ with radius r is given by 
\begin{equation}
(\cos\theta^{\nu_i}_r)_{\rm{HT}}=\frac{R^3_{\nu_i}r^2\Big(1-\frac{2M}{r}+\frac{2J^2}{r^4}\Big)^\frac{1}{2}}{2J(r^3-R_{\nu_i}^3)+R_{\nu_i}^2r^3\Big(1-\frac{2M}{R_{\nu_i}}+\frac{2J^2}{R_{\nu_i}^4}\Big)^\frac{1}{2}},
\label{eq:21}
\end{equation}
where $i=e, \mu, \tau$. In the Newtonian limit, Eq.\ref{eq:21} becomes $(\cos\theta^{\nu_i}_r)_{\rm{Newt}}=\frac{R_{\nu_i}}{r}$, and in the non rotating Schwarzschild limit Eq.\ref{eq:21} becomes $(\cos\theta^{\nu_i}_r)_{\rm{Sch}}=\frac{R_{\nu_i}}{r}\sqrt{\frac{1-\frac{2M}{r}}{1-\frac{2M}{R_{\nu_i}}}}$.

Puting the expressions of $\dot{t}$ and $\dot{\phi}$ (Eq.\ref{eq:15}) in Eq.\ref{eq:16} we obtain 
\begin{equation}
 \frac{E^2}{2}=\frac{\dot{r}^2}{2}+V_{\rm{eff}},
\end{equation}
 where the effective potential $V_{\rm{eff}}$ defining the trajectory of a massless particle for a slowly rotating neutron star system is given as 
\begin{equation}
V_{\rm{eff}}=\frac{L^2}{2r^2}\Big(1-\frac{2M}{r}-\frac{2J^2}{r^4}\Big)+\frac{2JL^2}{b_Ir^3}.
\label{eq:22}
\end{equation}
To find the neutrinosphere radius which corresponds to the last stable circular orbit for the neutrinos, we have to impose $\frac{dV_{eff}}{dr}=0$. Then from Eq.\ref{eq:22} we obtain (Note that our expressions Eq.\ref{eq:23} and Eq.\ref{eq:24} are slightly different from those given in \cite{Prasanna:2001ie} due to some typographical errors.)
\begin{equation}
r^4-r^3\Big(3M-\frac{6J}{b_I}\Big)-6J^2=0.
\label{eq:23}
\end{equation}
If there is no rotation (Schwarzschild solution), then we get the neutrinosphere orbit at a radius $=3M$. In that case, the massless neutrino which has a radius $<3M$ is gravitationally bound. However, if we solve Eq.\ref{eq:23} for the rotating case, we can find at least one real root for which the neutrinosphere radius is $<3M$. If the neutrinos are emitted tangentially from the neutrinosphere surface, then from Eq.\ref{eq:20} we can also write, 
\begin{equation}
r^6-b_I^2r^4+2Mb_I^2r^3+2J^2b_I^2-4b_IJr^3=0.
\label{eq:24}
\end{equation}
\begin{table}[h]
\centering
\setlength{\tabcolsep}{15pt} 
\renewcommand{\arraystretch}{1.5} 
\begin{tabular}{ lccc  }
 
 \hline
$\frac{J}{M^2}$ &  $\frac{R_{EH}}{M}$ &  $\frac{R_\nu}{M}$ &  $\frac{b_I}{M}$\\
 \hline
0.1  & 2.002  & 2.882 &4.990 \\
0.2 & 2.009  &2.759 &4.770 \\
 0.3 & 2.022  & 2.632 &4.534\\
0.4 & 2.038 & 2.500 & 4.278 \\
0.5 & 2.057 & 2.363 & 3.998\\ 
 \hline
\end{tabular}
\caption{ \label{grf} Summary of the event horizon $(\frac{R_{EH}}{M})$, neutrinosphere radius $(\frac{R_\nu}{M})$, and impact parameter $(\frac{b_I}{M})$ for different values of $\frac{J}{M^2}$.}
\end{table}
We can calculate the neutrinosphere radius and the impact parameter by solving Eq.\ref{eq:23} and Eq.\ref{eq:24} simultaneously for different values of $\frac{J}{M^2}$. The event horizon is calculated using $1-\frac{2M}{R_{EH}}-\frac{2J^2}{R_{EH}^4}=0$. We are interested in the domain $r>R_\nu$ for the energy deposition rate. In TABLE \ref{grf} we have summarized the event horizons, neutrinosphere radius, and impact parameter for different values of $\frac{J}{M^2}$.

Now we compute the angular integration factor $\Theta(r)$ which appears in Eq.\ref{eq:9}. Choosing $d\Omega=d\mu d\phi$, where $\mu=\sin\theta$ and $\Omega=(\mu, \sqrt{1-\mu^2}\cos\phi, \sqrt{1-\mu^2}\sin\phi)$, we can write
\begin{equation}
\Theta(r)=4\pi^2\int^1_{x_{\nu_i}}\int^1_{x_{\nu_i}}\Big[1-2\mu_\nu \mu_{\bar{\nu}}+\mu^2_\nu\mu^2_{\bar{\nu}}+\frac{1}{2}(1-\mu^2_\nu)(1-\mu^2_{\bar{\nu}})\Big] d\mu_\nu d\mu_{\bar{\nu}}.
\label{eq:25}
\end{equation}
The solution of Eq.\ref{eq:25} in HT background is found as 
\begin{equation}
\Theta^{\rm{HT}}(r)=\frac{2\pi^2}{3}(1-x^{\rm{HT}}_{\nu_i})^4({x^{\rm{HT}}}^2_{\nu_i}+4x^{\rm{HT}}_{\nu_i}+5),
\label{eq:26}
\end{equation}
where,
\begin{equation}
x^{\rm{HT}}_{\nu_i}=(\sin\theta^{\nu_i}_r)^{\rm{HT}}=\Bigg[1-\frac{R^6_{\nu_i}r^4\Big(1-\frac{2M}{r}+\frac{2J^2}{r^4}\Big)}{\Bigg(2J(r^3-R^3_{\nu_i})+R^2_{\nu_i}r^3\Big(1-\frac{2M}{R_{\nu_i}}+\frac{2J^2}{R^4_{\nu_i}}\Big)^\frac{1}{2}\Bigg)^2}\Bigg]^\frac{1}{2}.
\label{eq:27}
\end{equation}
The temperature of the free streaming neutrinos at a radius $r$ is related to the temperature of the neutrino at the neutrinosphere by the gravitational redshift. The neutrino temperature varies linearly with the redshift as
\begin{equation}
T^{\rm{HT}}_{\nu_i}(r)=\sqrt{\frac{1-\frac{2M}{R_{\nu_i}}-\frac{2J^2}{R_{\nu_i}^4}}{1-\frac{2M}{r}-\frac{2J^2}{r^4}}}T_{\nu_i}(R_{\nu_i}),
\label{eq:28}
\end{equation}
whereas the observable quantity, the luminosity varies quadratically with the redshift as 
\begin{equation}
L^{\rm{HT}}_{\rm{obs}}=\Big(1-\frac{2M}{R_{\nu_i}}-\frac{2J^2}{R_{\nu_i}^4}\Big)L_{\nu_i}(R_{\nu_i}).
\label{eq:29}
\end{equation}
The neutrino luminosity for each $\nu_i$ species for a blackbody neutrino gas can also be written as
\begin{equation}
L_{\nu_i}(R_{\nu_i})=4\pi R^2_{\nu_i} \frac{7}{16}aT_{\nu_i}^4(R_{\nu_i}),
\label{eq:30}
\end{equation}
where $a=0.663$ is the radiation constant in natural units.

Using Eq.\ref{eq:26}, Eq.\ref{eq:28}, Eq.\ref{eq:29}, Eq.\ref{eq:30} we can calcualte
\begin{equation}
{T^{\rm{HT}}}^9_{\nu_i}(r)\Theta^{\rm{HT}}_{\nu_i}(r)=\frac{\Big(1-\frac{2M}{R_{\nu_i}}-\frac{2J^2}{R^4_{\nu_i}}\Big)^\frac{9}{4}}{\Big(1-\frac{2M}{r}-\frac{2J^2}{r^4}\Big)^\frac{9}{2}}\Big(\frac{7}{4}\pi a\Big)^{-\frac{9}{4}}R_{\nu_i}^{-\frac{9}{2}}{L^{\rm{HT}}}^{\frac{9}{4}}_{\rm{obs}}\times \frac{2\pi^2}{3}(1-x^{\rm{HT}}_{\nu_i})^4({x^{\rm{HT}}}^2_{\nu_i}+4x^{\rm{HT}}_{\nu_i}+5).
\label{eq:31}
\end{equation}
The energy deposition rate is enhanced with increasing $T^9_{\nu_i}(r)\Theta_{\nu_i}(r)$ which has distinct values in different background spacetimes.

Note that, the limit $J=0$ ($J=0$, $M=0$) corresponds to Schwarzschild (Newtonian) background.
\subsection{Born-Infeld generalization of Reissner-Nordstrom solution}
In the following, we will discuss the energy deposition for the Born-Infeld Reissner-Nordstrom (BIRN) gravity solution. The reason is that in this background the energy deposition matches quite well with the observed maximum GRB energy for certain values of modified gravity parameters.

In the Born-Infeld model, the nonlinear electromagnetic generalization of the Reissner-Nordstrom solution is defined by the metric as \cite{Breton:2002td}
\begin{equation}
ds^2=-f(r)dt^2+{f(r)}^{-1}dr^2+r^2d\Omega^2,
\label{rn1}
\end{equation}
where 
\begin{equation}
f(r)=1-\frac{2M}{r}+\frac{2}{3}b^2r^2\Big(1-\sqrt{1+\frac{Q^2}{b^2r^4}}\Big)+\frac{2Q^2}{3r}\sqrt{\frac{b}{Q}}F\Big(\arccos\Big(\frac{br^2/Q-1}{br^2/Q+1}\Big),\frac{1}{\sqrt{2}}\Big).
\label{rn2}
\end{equation}
Here, $b$ denotes the Born-Infeld parameter defined as the magnitude of the electric field at $r=0$, $Q$ denotes the electric charge, and $F$ denotes the Legendre's elliptic function of the first kind given as $F(\beta,\alpha)=\int^\beta_0 (1-\alpha\sin^2\theta)^{-\frac{1}{2}}$.
\begin{figure}[!htbp]
\centering
\subfigure[$f(\frac{r}{M})$ vs. $\frac{r}{M}$, with different $Q$ and $b$.]{\includegraphics[width=8cm]{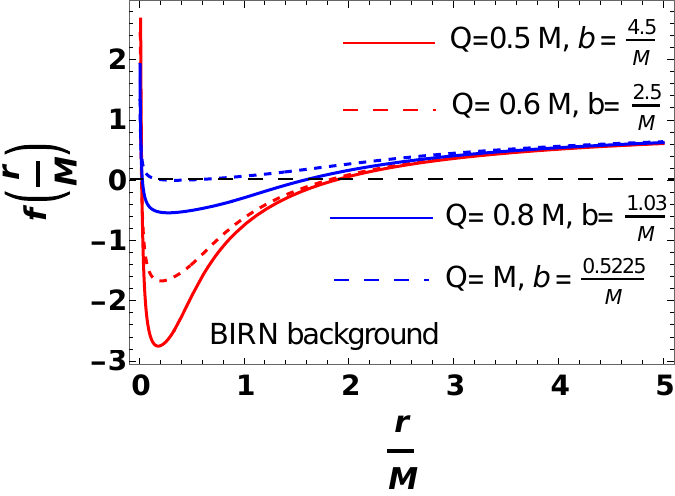}\label{plotrina}}
\subfigure[$f(\frac{r}{M})$ vs. $\frac{r}{M}$, with fixed $Q$ and different $b$. ]{\includegraphics[width=8cm]{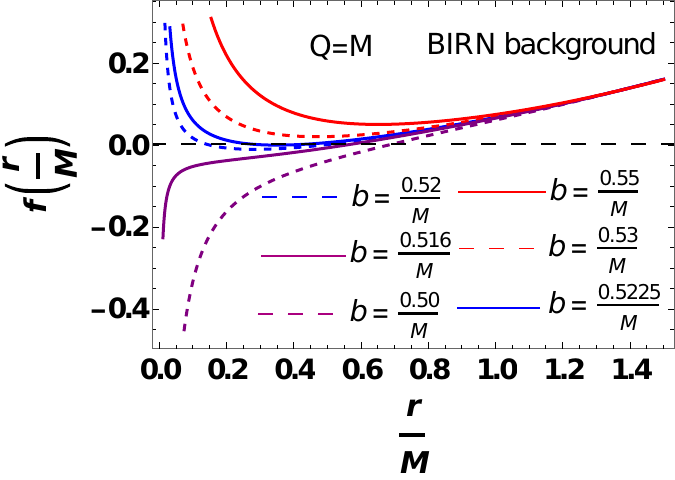}\label{plotrinb}}
\caption{(a)Variation of $f(\frac{r}{M})$ with respect to $\frac{r}{M}$ for different values of charge and Born-Infeld parameter in BIRN background. (b)Variation of $f(\frac{r}{M})$ with respect to $\frac{r}{M}$ for different values of Born-Infeld parameter with $Q=M$ in BIRN background.}
\label{plotrin}
\end{figure}

In Fig.\ref{plotrina} we have shown the variation of $f(\frac{r}{M})$ with respect to $\frac{r}{M}$ for different values of $Q$ and $b$. The location of the horizon corresponds to the position for which $f(\frac{r}{M})=0$. The values of $b$ are chosen in such a way that as $r\rightarrow 0$, $f(\frac{r}{M})\rightarrow +\infty$ which corresponds to the soliton like behaviour. For $0.5M\lesssim Q\lesssim 0.9M$, $f(\frac{r}{M})$ has two zeros. For $Q=0.5M$, the metric has a soliton like behavoiur for $b\gtrsim\frac{4.2}{M}$.

In Fig.\ref{plotrinb} we consider the special case $Q=M$ which corresponds to the extreme case in black hole terminology. In the extreme case with $b=0.5225/M$ (blue solid line), the metric Eq.\ref{rn1} has one horizon and goes to $+\infty$ near $r=0$. For $b\gtrsim 0.53/M$ (red dashed line), the metric has a naked singularity, and for $b\leq \frac{0.516}{M}$ (solid purple line), the metric behaves as a Schwarzschild spacetime. 

In BIRN background we can write
\begin{equation}
\begin{split}
{T^{\rm{BIRN}}}^9_{\nu_i}(r)\Theta^{\rm{BIRN}}_{\nu_i}(r)=\frac{\Big(1-\frac{2M}{R}+\frac{2}{3}b^2R^2\Big(1-\sqrt{1+\frac{Q^2}{b^2R^4}}\Big)+\frac{2Q^2}{3R}\sqrt{\frac{b}{Q}}F\Big(\arccos\Big(\frac{bR^2/Q-1}{bR^2/Q+1}\Big),\frac{1}{\sqrt{2}}\Big)\Big)^\frac{9}{4}}{\Big(1-\frac{2M}{r}+\frac{2}{3}b^2r^2\Big(1-\sqrt{1+\frac{Q^2}{b^2r^4}}\Big)+\frac{2Q^2}{3r}\sqrt{\frac{b}{Q}}F\Big(\arccos\Big(\frac{br^2/Q-1}{br^2/Q+1}\Big),\frac{1}{\sqrt{2}}\Big)\Big)^\frac{9}{2}}\times\\
\Big(\frac{7}{4}\pi a\Big)^{-\frac{9}{4}}R^{-\frac{9}{2}}_{\nu_i}{L^{\rm{BIRN}}}^{\frac{9}{4}}_{\rm{obs}}\times \frac{2\pi^2}{3}(1-x^{\rm{BIRN}}_{\nu_i})^4({x^{\rm{BIRN}}}^2_{\nu_i}+4x^{\rm{BIRN}}_{\nu_i}+5),
\end{split}
\label{rn3}
\end{equation}
where $L^{\rm{BIRN}}_{\rm{obs}}$ in BIRN background is 
\begin{equation}
L^{\rm{BIRN}}_{\rm{obs}}=\Big(1-\frac{2M}{R_{\nu_i}}+\frac{2}{3}b^2R_{\nu_i}^2\Big(1-\sqrt{1+\frac{Q^2}{b^2R_{\nu_i}^4}}\Big)+\frac{2Q^2}{3R_{\nu_i}}\sqrt{\frac{b}{Q}}F\Big(\arccos\Big(\frac{bR_{\nu_i}^2/Q-1}{bR_{\nu_i}^2/Q+1}\Big),\frac{1}{\sqrt{2}}\Big)\Big)L_{\nu_i}(R_{\nu_i}),
\end{equation}
and $x_{\nu_i}$ is given as
\begin{equation}
x^{\rm{BIRN}}_{\nu_i}=\Bigg[1-\frac{R^2}{r^2}\Bigg(\frac{1-\frac{2M}{r}+\frac{2}{3}b^2r^2\Big(1-\sqrt{1+\frac{Q^2}{b^2r^4}}\Big)+\frac{2Q^2}{3r}\sqrt{\frac{b}{Q}}F\Big(\arccos\Big(\frac{br^2/Q-1}{br^2/Q+1}\Big),\frac{1}{\sqrt{2}}\Big)}{1-\frac{2M}{R}+\frac{2}{3}b^2R^2\Big(1-\sqrt{1+\frac{Q^2}{b^2R^4}}\Big)+\frac{2Q^2}{3R}\sqrt{\frac{b}{Q}}F\Big(\arccos\Big(\frac{bR^2/Q-1}{bR^2/Q+1}\Big),\frac{1}{\sqrt{2}}\Big)}\Bigg)\Bigg]^\frac{1}{2}.
\label{rn4}
\end{equation}
In the limit, $b\rightarrow \infty$, the metric turns into the linear Einstein-Maxwell Reissner-Nordstrom solution. If we further put $Q\rightarrow 0$, we obtain the Schwarzschild solution.
\subsection{Quintessence background}
Quintessence is a dynamical, time dependent, and spatially inhomogeneous scalar field which was first introduced as an alternative of cosmological constant $(\Lambda)$ to explain the accelerated expansion of the universe \cite{Caldwell:1997ii}. The quintessence field around a massive gravitating object can deform the spacetime. The spacetime geometry in presence of quintessence is parametrized by two quantities, the equation of state $\omega$ which can take values $-1<\omega<-\frac{1}{3}$ and the quintessence parameter $c$ \cite{Kiselev:2002dx}. The black hole solution surrounded by a quintessence field is governed by the metric as given in Eq.\ref{rn1} with different expressions of $f(r)$ as  
\begin{equation}
f(r)=1-\frac{2M}{r}-\frac{c}{r^{3\omega+1}}.
\label{cw}
\end{equation} 
Note that $c=\frac{\Lambda}{3}$ and $\omega=-1$ corresponds to the cosmological constant scenario.
\begin{figure}[!htbp]
\centering
\includegraphics[width=8cm]{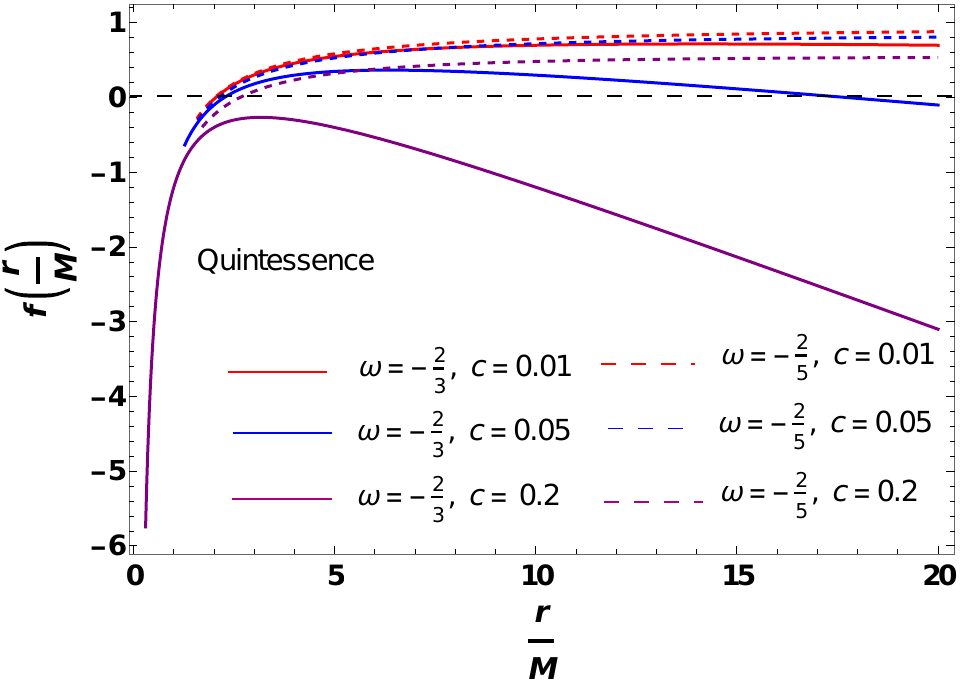}
\caption{Variation of $f(\frac{r}{M})$ with respect to $\frac{r}{M}$ for different values of $\omega$ and $c$ for a black hole solution surrounded by a quintessence field.}
\label{plotquint}
\end{figure}
The location of the event horizon is governed by the equation $f(\frac{r}{M})=0$. In Fig.\ref{plotquint} we have shown the variation of $f(\frac{r}{M})$ with $\frac{r}{M}$ for fixed values of $\omega=-\frac{2}{3}$, (solid lines) $-\frac{2}{5}$ (dashed lines) and different values of $c= 0.01$ (red), $0.05$ (blue), $0.2$ (purple). With increasing the value of $c$, the model approaches from a black hole solution to a naked singularity. The solid purple line corresponds to the naked singularity with $\omega=-\frac{2}{3}$ and $c=0.2$. The value of $f(\frac{r}{M})$ increases with increasing $\omega$. In the following, we will consider the black hole solution.  

For the black hole solution in presence of a quintessence field, Eq.\ref{eq:31} becomes
\begin{equation}
\begin{split}
{T^{\rm{Quint}}}^9_{\nu_i}(r)\Theta^{\rm{Quint}}_{\nu_i}(r)=\frac{(1-\frac{2M}{R}-\frac{c}{R^{3\omega+1}})^{\frac{9}{4}}}{(1-\frac{2M}{r}-\frac{c}{r^{3\omega+1}})^{\frac{9}{2}}}\Big(\frac{7}{4}\pi a\Big)^{-\frac{9}{4}}R^{-\frac{9}{2}}_{\nu_i}{L^{\rm{Quint}}}^{\frac{9}{4}}_{\rm{obs}}\times \frac{2\pi^2}{3}(1-x^{\rm{Quint}}_{\nu_i})^4\times \\
({x^{\rm{Quint}}}^2_{\nu_i}+4x^{\rm{Quint}}_{\nu_i}+5),
\end{split}
\end{equation}
where $L^{\rm{Quint}}_{\rm{obs}}$ in Quintessence background is
\begin{equation}
L^{\rm{Quint}}_{\rm{obs}}=\Big(1-\frac{2M}{R_{\nu_i}}-\frac{c}{R_{\nu_i}^{3\omega+1}}\Big)L_{\nu_i}(R_{\nu_i}),
\end{equation}
and $x^{\rm{Quint}}_{\nu_i}$ is
\begin{equation}
x^{\rm{Quint}}_{\nu_i}=\Big[1-\frac{R^2}{r^2}\Big(\frac{1-\frac{2M}{r}-\frac{c}{r^{3\omega+1}}}{1-\frac{2M}{R}-\frac{c}{R^{3\omega+1}}}\Big)\Big]^\frac{1}{2}.
\label{a1quint}
\end{equation}
Note that $c=0$ corresponds to the Schwarzschild solution.
\section{$Z^\prime$ contribution to neutrino heating in different spacetime background}\label{sec4}
In this section, we obtain expressions for the rate of energy deposition due to SM+$Z^\prime$ mediated neutrino heating in different spacetime backgrounds.
\subsection{Hartle-Thorne background}
The total amount of energy deposition in the HT background due to neutrino heating in the region beyond the neutrinosphere is given as
\begin{equation}
\dot{Q}^{\rm{HT}}_{\nu_i}=\int^\infty_{R_{\nu_i}}\dot{q}^{\rm{HT}}_{\nu_i}\frac{4\pi r^2dr}{\sqrt{1-\frac{2M}{r}-\frac{2J^2}{r^4}}}
\label{eq:32}
\end{equation}
The total energy deposition by all the neutrino species is $\dot{Q}^{\rm{HT}}_{\nu_e}+\dot{Q}^{\rm{HT}}_{\nu_{\mu,\tau}}$. If we only include the $W$ and $Z$ mediated diagrams \textit{i.e}, only the SM contribution for the process $\nu\overline{\nu}\rightarrow e^+e^-$ and denote $D=1\pm 4\sin^2\theta_w+8\sin^4\theta_w$, where $+$ sign is for $\nu_e\overline{\nu_e}$ pair, and $-$ sign is for $\nu_\mu\overline{\nu_\mu}$ and $\nu_\tau\overline{\nu_\tau}$ pairs then the total rate of energy deposition in HT spacetime background becomes
\begin{equation}
 \dot{Q}^{\rm{HT}}_{51}=1.09\times 10^{-5}F\Big(\frac{M}{R},\frac{J}{R^2}\Big)D{L^{\rm{HT}}}^{9/4}_{51}R_6^{-3/2},
 \label{eq:40}
\end{equation}
where $\dot{Q}_{51}=\frac{\dot{Q}}{10^{51}\hspace{0.1cm}\rm{erg/sec}}$, $L_{51}=\frac{L_{\rm{obs}}}{10^{51}\hspace{0.1cm}\rm{erg/sec}}$, $R_6=\frac{R}{10\hspace{0.1cm}\rm{km}}$, and the enhancement function 
\begin{equation}
 F\Big(\frac{M}{R},\frac{J}{R^2}\Big)=3\Big(1-\frac{2M}{R}-\frac{2J^2}{R^4}\Big)^{9/4}\int ^\infty_1 \frac{y^2dy}{\Big(1-\frac{2M}{yR}-\frac{2J^2}{(yR)^4}\Big)^5}(1-x^{\rm{HT}})^4({x^2}^{\rm{HT}}+4x^{\rm{HT}}+5),
 \label{eq:41}
\end{equation}
where we assume $R_{\nu_e}\approx R_{\nu_\mu,\nu_\tau}=R$, $r=yR$, and, $x_\nu^{\rm{HT}}=x^{\rm{HT}}$ is given by Eq \ref{eq:27}. Eq.\ref{eq:40} is for a single neutrino flavor and thus one needs to calculate the total energy deposition contributed by all three flavors. 

Putting $J=0$ in Eq.\ref{eq:41} gives the function $F$ for the Schwarzschild background with $x^{\rm{Sch}}=\Big[1-\frac{R^2}{r^2}\frac{(1-\frac{2M}{r})}{(1-\frac{2M}{R})}\Big]^\frac{1}{2}$ For $M\rightarrow 0$ and $J\rightarrow 0$, the above function becomes $F(0)=1$ corresponding to the Newtonian background with $x^{\rm{Newt}}=\Big(1-\frac{R^2}{r^2}\Big)^{\frac{1}{2}}$.
\begin{table}[h]
\centering
\setlength{\tabcolsep}{15pt} 
\renewcommand{\arraystretch}{1.5} 
\begin{tabular}{ lccc  }

 \hline
$\rm{Background}$ & $\frac{J}{M^2}$ &  $\frac{R}{M}$ &  $\dot{Q}~\rm{(erg/s)}$\\
 \hline
Newtonian  & $0$  & $0$ & $1.5\times 10^{50}$ \\
Schwarzschild & $0$  & $3$ &$4.3\times 10^{51}$ \\
Schwarzschild & $0$  & $5$ &$6.4\times 10^{50}$\\
Hartle-Thorne & $0.1$ & $3$ & $3.6\times 10^{51}$ \\
Hartle-Thorne & $0.8$ & $3$ & $1.7\times 10^{51}$\\ 
Hartle-Thorne & $0.1$ & $5$ & $6.2\times 10^{50}$ \\
Hartle-Thorne & $0.8$ & $5$ & $5.0\times 10^{50}$\\ 
 \hline
\end{tabular}
\caption{ \label{grfa} Rate of energy emission due to neutrino pair annihilation for short GRB in Newtonian, Schwarzschild, and Hartle-Thorne background. Here we only consider the SM processes.}
\end{table}

In TABLE \ref{grfa} we present the maximum energy deposited in Newtonian, Schwarzschild, and Hartle-Thorne backgrounds due to the neutrino pair annihilation process in SM. Here we have used Eq.\ref{eq:40} for the Hartle-Thorne background and the equivalent expressions for Schwarzschild and Newtonian backgrounds by taking $J=0$ and $J=M=0$ in Eq.\ref{eq:40} respectively. We also consider that the neutrinos are emitted from the neutrinosphere, the value of $D$ is $1.23$, the neutrino luminosity at infinity is $\sim 10^{53}~\rm{erg}/s$, and $R=20\hspace{0.1cm}\rm{km}$. We infer from TABLE \ref{grfa} that the observed maximum GRB energy cannot be explained by modifying the background spacetime with Schwarzschild geometry. If we include rotation in the background spacetime (Hartle-Thorne), the rate of energy deposition decreases as compared to the Schwarzschild case.

If the $Z^\prime$ mediated process is included then Eq.\ref{eq:40} will be modified. However, Eq.\ref{eq:41} remains unchanged for any particular spacetime mentioned above, as it does not depend on the BSM physics. In the following, we obtain expressions for the rate of energy deposition in the Hartle-Thorne background including the $Z^\prime$ mediated neutrino pair annihilation process.
Using Eq.\ref{eq:8}, Eq.\ref{eq:10}, and Eq.\ref{eq:31}, we can write 
the energy deposition rate for the electron neutrino in HT metric as 
\begin{equation}
\begin{split}
\dot{Q}^{\rm{HT}}_{\nu_e}=\frac{28\pi^7}{(2\pi)^6}k^9\zeta(5)\times\Big[\frac{G^2_F}{3\pi}(1+4\sin^2\theta_W+8\sin^4\theta_W)+\frac{4g^{\prime^4}}{6\pi M_{Z^\prime}^4}+\\
\frac{4 G_F g^{\prime^2}}{3\sqrt{2}\pi M^2_{Z^\prime}}\Big(-\frac{1}{2}+2\sin^2\theta_W\Big)+\frac{4G_Fg^{\prime^2}}{3\sqrt{2}\pi M_{Z^\prime}^2 }\Big]\Big(1-\frac{2M}{R_{\nu_e}}-\frac{2J^2}{R^4_{\nu_e}}\Big)^\frac{9}{4}\Big(\frac{7\pi a}{4}\Big)^{-\frac{9}{4}}R_{\nu_e}^{-\frac{3}{2}}L^{\frac{9}{4}}_{\rm{obs}}\\
\int ^\infty_1 \frac{y^2_{\nu_e}dy_{\nu_e}}{\Big(1-\frac{2M}{y_{\nu_e}R_{\nu_e}}-\frac{2J^2}{(y_{\nu_e}R_{\nu_e})^4}\Big)^5}(1-x^{\rm{HT}}_{\nu_e})^4(x{^2_{\nu_e}}^{\rm{HT}}+4x^{\rm{HT}}_{\nu_e}+5),
\end{split}
\label{eq:34}
\end{equation}
where $y_{\nu_i}=\frac{r}{R_{\nu_i}}$ and $x^{\rm{HT}}_{\nu_e}$ is given by Eq.\ref{eq:27} for $\nu_e$. Similarly, the energy deposition rates for $\nu_\mu$ and $\nu_\tau$ in the HT background become 
\begin{equation}
\begin{split}
\dot{Q}^{\rm{HT}}_{\nu_{\mu,\tau}}=\frac{28\pi^7}{(2\pi)^6}k^9\zeta(5)\times\Big[\frac{G^2_F}{3\pi}(1-4\sin^2\theta_W+8\sin^4\theta_W)+\frac{4g^{\prime^4}}{6\pi M_{Z^\prime}^4}+\\
\frac{4 G_F g^{\prime^2}}{3\sqrt{2}\pi M^2_{Z^\prime}}\Big(-\frac{1}{2}+2\sin^2\theta_W\Big)\Big]\Big(1-\frac{2M}{R_{{\nu_{\mu,\tau}}}}-\frac{2J^2}{R^4_{{\nu_{\mu,\tau}}}}\Big)^\frac{9}{4}\Big(\frac{7\pi a}{4}\Big)^{-\frac{9}{4}}R_{{\nu_{\mu,\tau}}}^{-\frac{3}{2}}L^{\frac{9}{4}}_{\rm{obs}}\\
\int ^\infty_1 \frac{y^2_{\nu_{\mu,\tau}}dy_{\nu_{\mu,\tau}}}{\Big(1-\frac{2M}{y_{\nu_{\mu,\tau}}R_{{\nu_{\mu,\tau}}}}-\frac{2J^2}{(y_{\nu_{\mu,\tau}}R_{{\nu_{\mu,\tau}}})^4}\Big)^5}(1-x^{\rm{HT}}_{{\nu_{\mu,\tau}}})^4(x{^2_{\nu_{{\mu,\tau}}}}^{\rm{HT}}+4x^{\rm{HT}}_{{\nu_{\mu,\tau}}}+5),
\end{split}
\label{eq:35}
\end{equation}
where $x^{\rm{HT}}_{\nu_{\mu,\tau}}$ is given by Eq.\ref{eq:27} for $\nu_{\mu,\tau}$.

From Eq.\ref{eq:34} and Eq.\ref{eq:35}, we can similarly obtain the energy deposition rates for three flavors of neutrinos in the Schwarzschild and the Newtonian background with proper choices of $x^{\rm{Sch}}_{\nu}$ and $x^{\rm{Newt}}_{\nu}$ by putting $J=0$ and $J=M=0$ in Eq.\ref{eq:27} respectively.
\subsection{Born-Infeld generalization of Reissner-Nordstrom solution}
If we only include the SM neutrino pair annihilation processes, then the total rate of energy deposition in the BIRN background becomes
\begin{equation}
\dot{Q}^{\rm{BIRN}}_{51}=1.09\times 10^{-5}F\Big(\frac{M}{R},\frac{Q}{bR^2}\Big)D {L^{\rm{BIRN}}}^{9/4}_{51}R^{-3/2}_6,
\end{equation}
where $F\Big(\frac{M}{R},\frac{Q}{bR^2}\Big)$ is given as 
\begin{equation}
F\Big(\frac{M}{R},\frac{Q}{bR^2}\Big)=3f(R)^{9/4}\int^\infty_1 \frac{y^2dy}{f(yR)^5}(1-x^{BIRN}_\nu)^4({x^2}^{BIRN}_\nu+4x^{BIRN}_\nu+5),
\end{equation}
where $f(r)$ is defined in Eq.\ref{rn2}. Putting all the values of the parameters as given above and choosing $\frac{R}{M}=2$, $Q=M$ and $b=\frac{0.30}{M}$ we obtain the total rate of energy deposition $\dot{Q}^{\rm{BIRN}}_{51}\sim 1.5\times 10^{52}~\rm{erg/s}$. This value matches quite well with the maximum GRB energy. 

We consider that the contribution of $Z^\prime$ in neutrino heating is limited to be no larger than the experimental uncertainty in the measurement of GRB energy.  
We can calculate the energy deposition rate for $\nu_e$ in BIRN background including the $Z^\prime$ mediated process as
\begin{equation}
\begin{split}
\dot{Q}^{\rm{BIRN}}_{\nu_e}=\frac{28\pi^7}{(2\pi)^6}k^9\zeta(5)\times\Big[\frac{G^2_F}{3\pi}(1+4\sin^2\theta_W+8\sin^4\theta_W)+\frac{4g^{\prime^4}}{6\pi M_{Z^\prime}^4}+\\
\frac{4 G_F g^{\prime^2}}{3\sqrt{2}\pi M^2_{Z^\prime}}\Big(-\frac{1}{2}+2\sin^2\theta_W\Big)+\frac{4G_Fg^{\prime^2}}{3\sqrt{2}\pi M_{Z^\prime}^2 }\Big]\Big(1-\frac{2M}{R_{\nu_e}}+\frac{2}{3}b^2R_{\nu_e}^2\Big(1-\sqrt{1+\frac{Q^2}{b^2R_{\nu_e}^4}}\Big)+\\
\frac{2Q^2}{3R_{\nu_e}}\sqrt{\frac{b}{Q}}F\Big(\arccos\Big(\frac{bR_{\nu_e}^2/Q-1}{bR_{\nu_e}^2/Q+1}\Big),\frac{1}{\sqrt{2}}\Big)\Big)^\frac{9}{4}\Big(\frac{7\pi a}{4}\Big)^{-\frac{9}{4}}R_{{\nu_e}}^{-\frac{3}{2}}L^{\frac{9}{4}}_{\rm{obs}}\times\\
\int^\infty_1 \frac{y^2_{\nu_e}dy_{\nu_e}}{\Big(1-\frac{2M}{y_{\nu_e}R_{\nu_e}}+\frac{2}{3}b^2y_{\nu_e}^2R_{\nu_e}^2\Big(1-\sqrt{1+\frac{Q^2}{b^2y_{\nu_e}^4R_{\nu_e}^4}}\Big)+\frac{2Q^2}{3y_{\nu_e}R_{\nu_e}}\sqrt{\frac{b}{Q}}F\Big(\arccos\Big(\frac{by_{\nu_e}^2R_{\nu_e}^2/Q-1}{by_{\nu_e}^2R_{\nu_e}^2/Q+1}\Big),\frac{1}{\sqrt{2}}\Big)\Big)^5}\times\\
(1-x^{\rm{BIRN}}_{\nu_e})^4(x{^2_{\nu_e}}^{\rm{BIRN}}+4x^{\rm{BIRN}}_{\nu_e}+5),
\end{split}
\label{birn1}
\end{equation}
where the expression of $x^{BIRN}_{\nu_i}$ is given in Eq.\ref{rn4}. 
Similarly, the energy deposition for $\nu_\mu$ and $\nu_\tau$ in BIRN background is same as Eq.\ref{birn1} with $y_{\nu_e}$, $R_{\nu_e}$ and $x_{\nu_e}$ will be replaced by $y_{\nu_{\mu,\tau}}$, $R_{\nu_{\mu,\tau}}$ and $x_{\nu_{\mu,\tau}}$ respectively. In addition, there will be a negative sign in front of $4\sin^2\theta_W$ and the last term in the square bracket should be omitted due to only contributions from $Z$ and $Z^\prime$ mediated diagrams.  
\subsection{Quintessence background} 
The total rate of energy deposition for the SM neutrino pair annihilation process in presence of a Quintessence background becomes 
\begin{equation}
\dot{Q}^{\rm{Quint}}_{51}=1.09\times 10^{-5}F\Big(\frac{M}{R},\frac{c}{R^{3\omega}+1}\Big)D{L^{\rm{Quint}}}^{9/4}_{51}R^{-3/2}_6,
\end{equation}
where $F\Big(\frac{M}{R},\frac{c}{R^{3\omega}+1}\Big)$ is given as 
\begin{equation}
F\Big(\frac{M}{R},\frac{c}{R^{3\omega}+1}\Big)=3f(R)^{9/4}\int^\infty_1 \frac{y^2dy}{f(yR)^5}(1-x^{Quint}_\nu)^4({x^2}^{Quint}_\nu+4x^{Quint}_\nu+5),
\end{equation}
where $f(r)$ is defined in Eq.\ref{cw}. For $\frac{R}{M}=3$, $\omega=-\frac{2}{3}$, and $c=2.487\times 10^{-2}$ we obtain the total rate of energy deposition $\dot{Q}^{\rm{Quint}}_{51}\sim 9.99\times 10^{51}~\rm{erg/s}$ which matches well with the maximum GRB energy. 

We consider that in quintessence background the contribution of $Z^\prime$ is limited to be no larger than the measurement uncertainty.
The energy deposition due to electron neutrino pair annihilation in quintessence background including the $Z^\prime$ mediated process is 
\begin{equation}
\begin{split}
\dot{Q}^{\rm{Quint}}_{\nu_e}=\frac{28\pi^7}{(2\pi)^6}k^9\zeta(5)\times\Big[\frac{G^2_F}{3\pi}(1+4\sin^2\theta_W+8\sin^4\theta_W)+\frac{4g^{\prime^4}}{6\pi M_{Z^\prime}^4}+\\
\frac{4 G_F g^{\prime^2}}{3\sqrt{2}\pi M^2_{Z^\prime}}\Big(-\frac{1}{2}+2\sin^2\theta_W\Big)+\frac{4G_Fg^{\prime^2}}{3\sqrt{2}\pi M_{Z^\prime}^2 }\Big]\Big(1-\frac{2M}{R_{\nu_e}}-\frac{c}{R_{\nu_e}^{3\omega+1}}\Big)^{\frac{9}{4}}\Big(\frac{7\pi a}{4}\Big)^{-\frac{9}{4}}R_{\nu_e}^{-\frac{3}{2}}L^{\frac{9}{4}}_{\rm{obs}}\\
\int ^\infty_1 \frac{y^2_{\nu_e}dy_{\nu_e}}{\Big(1-\frac{2M}{y_{\nu_e}R_{\nu_e}}-\frac{c}{(y_{\nu_e}R_{\nu_e})^{3\omega+1}}\Big)^5}(1-x^{\rm{Quint}}_{\nu_e})^4(x{^2_{\nu_e}}^{\rm{Quint}}+4x^{\rm{Quint}}_{\nu_e}+5),
\end{split}
\end{equation} 
where $x^{\rm{Quint}}_{\nu_e}$ is given br Eq.\ref{a1quint}. Similarly, the energy deposition due to muon and tau neutrino annihilation is governed by
\begin{equation}
\begin{split}
\dot{Q}^{\rm{Quint}}_{\nu_{\mu,\tau}}=\frac{28\pi^7}{(2\pi)^6}k^9\zeta(5)\times\Big[\frac{G^2_F}{3\pi}(1-4\sin^2\theta_W+8\sin^4\theta_W)+\frac{4g^{\prime^4}}{6\pi M_{Z^\prime}^4}+\\
\frac{4 G_F g^{\prime^2}}{3\sqrt{2}\pi M^2_{Z^\prime}}\Big(-\frac{1}{2}+2\sin^2\theta_W\Big)\Big]\Big(1-\frac{2M}{R_{{\nu_{\mu,\tau}}}}-\frac{c}{R^{3\omega+1}_{{\nu_{\mu,\tau}}}}\Big)^\frac{9}{4}\Big(\frac{7\pi a}{4}\Big)^{-\frac{9}{4}}R_{{\nu_{\mu,\tau}}}^{-\frac{3}{2}}L^{\frac{9}{4}}_{\rm{obs}}\\
\int ^\infty_1 \frac{y^2_{\nu_{\mu,\tau}}dy_{\nu_{\mu,\tau}}}{\Big(1-\frac{2M}{y_{\nu_{\mu,\tau}}R_{{\nu_{\mu,\tau}}}}-\frac{c}{(y_{\nu_{\mu,\tau}}R_{{\nu_{\mu,\tau}}})^{3\omega+1}}\Big)^5}(1-x^{\rm{Quint}}_{{\nu_{\mu,\tau}}})^4(x{^2_{\nu_{{\mu,\tau}}}}^{\rm{Quint}}+4x^{\rm{Quint}}_{{\nu_{\mu,\tau}}}+5),
\end{split}
\end{equation}
where $x^{\rm{Quint}}_{{\nu_{\mu,\tau}}}$ is given by Eq.\ref{a1quint} for $\nu_{\mu,\tau}$.
\section{Quantitative estimates of neurino heating due to $Z^\prime$ contribution in diffferent spacetime backgrounds}\label{sec5}
In this section, we study the effect of $Z^\prime$ mediated contribution to the pair annihilation of neutrinos $(\nu\overline{\nu})$ and calculate the energy deposited in GRB for Newtonian, Schwarzschild, Hartle-Thorne, Born-Infeld Reissner-Nordstrom, and Quintessence backgrounds.
\subsection{Newtonian Background}\label{subsec1}
\begin{figure}[h]
\includegraphics[height=8cm]{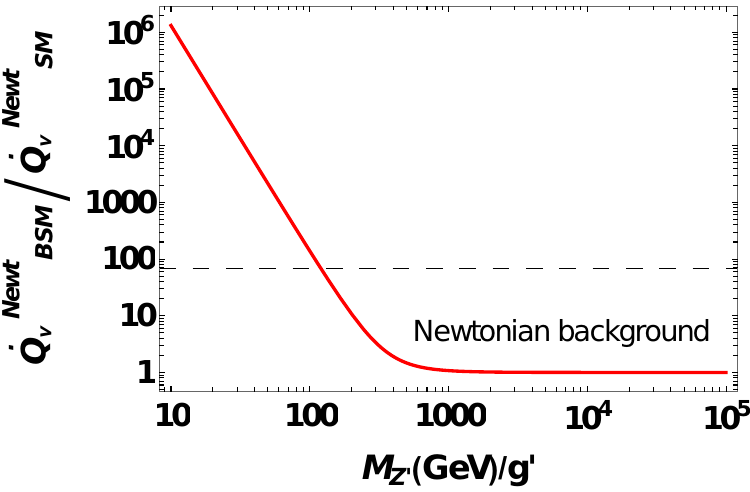}
\caption{ Variation of the ratio of rate of energy depositions in BSM and SM processes with respect to $\frac{M_{Z^\prime}}{g^\prime}$ in Newtonian background.} 
\label{plot1}
\end{figure} 
In Newtonian background, the SM contribution to $\nu\overline{\nu}\rightarrow e^+e^-$ process gives the value of the energy deposition as $\sim 1.5\times 10^{50}\hspace{0.1cm}\rm{erg}$ whereas the maximum energy in a GRB is $10^{52}\hspace{0.1cm}\rm{erg}$. We investigate to what extent the inclusion of $Z^\prime$ mediated process can energize a GRB. In Fig.\ref{plot1}, the red line denotes the variation of the ratio of the rate of energy depositions in BSM to that in SM with respect to $\frac{M_{Z^\prime}}{g^\prime}$. The figure shows that the ratio decreases with increasing $\frac{M_{Z^\prime}}{g^\prime}$. As $\frac{M_{Z^\prime}}{g^\prime}\rightarrow \infty$ (here it happens at $\frac{M_{Z^\prime}}{g^\prime}\gtrsim 10^3~\rm{GeV}$) the BSM effect goes away as expected and the ratio becomes unity. For smaller values of $\frac{M_{Z^\prime}}{g^\prime}$ there can be a greater enhancement in the energy deposition compared to the SM contribution. The black dashed line denotes the enhancement in energy required with respect to that of the SM+Newtonian background to explain the maximum energy in GRB. 

\subsection{Schwarzschild Background}\label{subsec2}
As discussed earlier, including the GR effects can enhance the rate of energy deposition via neutrino annihilation \cite{Salmonson:1999es} compared to the Newtonian calculation. In this section, we explore the effect of BSM physics if the background spacetime is Schwarzschild. 
\begin{figure}[h]
\includegraphics[height=8cm]{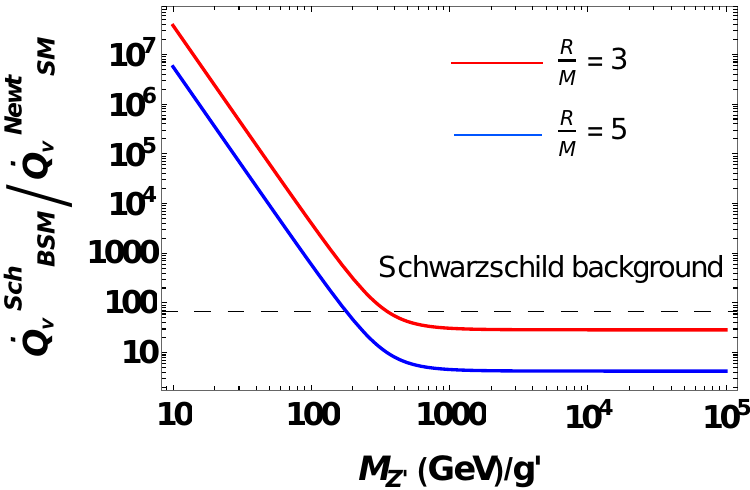}
\caption{Variation of the ratio of rate of energy depositions in BSM+Schwarzschild and SM+Newtonian processes with respect to $\frac{M_{Z^\prime}}{g^\prime}$. We have shown the variations for $\frac{R}{M}=3$ (red) and $\frac{R}{M}=5$ (blue).} 
\label{plot3}
\end{figure}

In Fig.\ref{plot3} we have shown the variation of the ratio of energy deposition in BSM+Schwarzschild to SM+Newtonian cases with respect to $\frac{M_{Z^\prime}}{g^\prime}$. We have shown the variations for $\frac{R}{M}=3$ (red) and $\frac{R}{M}=5$ (blue) which are typical values for merging neutron stars. From the figure, it is seen that the ratio decreases with increasing values of $\frac{R}{M}$ and $\frac{M_{Z^\prime}}{g^\prime}$.
For a very large value of $\frac{M_{Z^\prime}}{g^\prime}(\gtrsim 10^3~\rm{GeV})$ the ratio becomes constant. Note that, in this limit the BSM effects are negligible and the enhancement in energy deposition obtained is $\sim 4-30$ (depending on the values of $\frac{R}{M}$) because of changing the spacetime geometry from Newtonian to Schwarzschild. The black dashed line denotes the enhancement in energy required with respect to that of the SM+Newtonian background to explain the maximum energy in GRB.  
\begin{figure}[h]
\includegraphics[height=8cm]{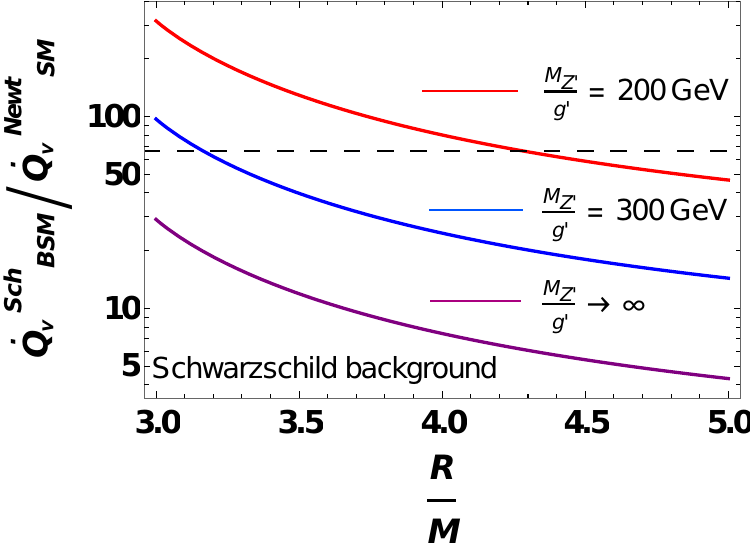}
\caption{Variation of the ratio of rate of energy depositions in BSM+Schwarzschild and SM+Newtonian processes with respect to $\frac{R}{M}$. We have shown the variations for $\frac{M_{Z^\prime}}{g^\prime}=200~\rm{GeV}$ (red), $\frac{M_{Z^\prime}}{g^\prime}=300~\rm{GeV}$ (blue), and $\frac{M_{Z^\prime}}{g^\prime}\rightarrow \infty$ (purple).}
\label{plot4}
\end{figure}

In Fig.\ref{plot4} we have shown the variation of the ratio of the rate of energy depositions in BSM+Schwarzschild and SM+Newtonian cases with respect to $\frac{R}{M}$. We have varied $\frac{R}{M}$ from $3$ to $5$ which is relevant for the merging neutron stars. We have shown the variations for $\frac{M_{Z^\prime}}{g^\prime}=200~\rm{GeV}$ (red), $\frac{M_{Z^\prime}}{g^\prime}=300~\rm{GeV}$ (blue), and $\frac{M_{Z^\prime}}{g^\prime}\rightarrow \infty$ (purple). The ratio decreases as we increase the values of $\frac{M_{Z^\prime}}{g^\prime}$ and $\frac{R}{M}$. This is expected since larger values of $M_{Z^\prime}/g^\prime$ tend towards the SM case and larger $\frac{R}{M}$ values correspond to the Newtonian case. The black dashed line denotes the enhancement in energy required with respect to that of the SM+Newtonian background to explain the maximum energy in GRB.  
If there is no BSM physics included, the  Schwarzschild background enhances the energy deposition by a factor $\sim 30$ over the Newtonian background (black) at $\frac{R}{M}=3$. For smaller values of $\frac{R}{M}$ and $\frac{M_{Z^\prime}}{g^\prime}$ there can be a greater enhancement in the energy deposition compared to the SM+Newtonian contribution to explain the maximum energy in GRBs.

\subsection{Hartle-Thorne Background}\label{subsec3}
The Hartle-Thorne metric which includes the rotation can also enhance the energy deposition in neutrino heating as compared to the Newtonian background.
\begin{figure}[!htbp]
\centering
\subfigure[${{\dot{Q}}^{\rm{HT}}}_{\rm{BSM}}$ vs. $\frac{M_{Z^\prime}}{g^\prime}$, fixed $\frac{R}{M}$]{\includegraphics[width=8cm]{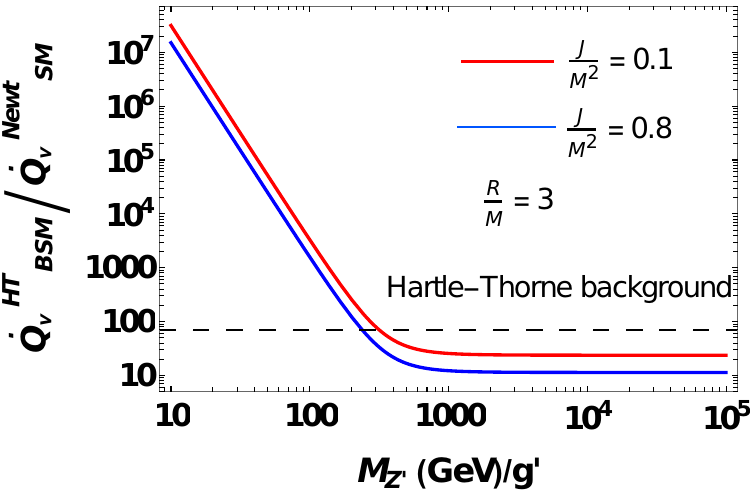}\label{plot6a}}
\subfigure[${{\dot{Q}}^{\rm{HT}}}_{\rm{BSM}}$ vs. $\frac{M_{Z^\prime}}{g^\prime}$, fixed $\frac{J}{M^2}$]{\includegraphics[width=8cm]{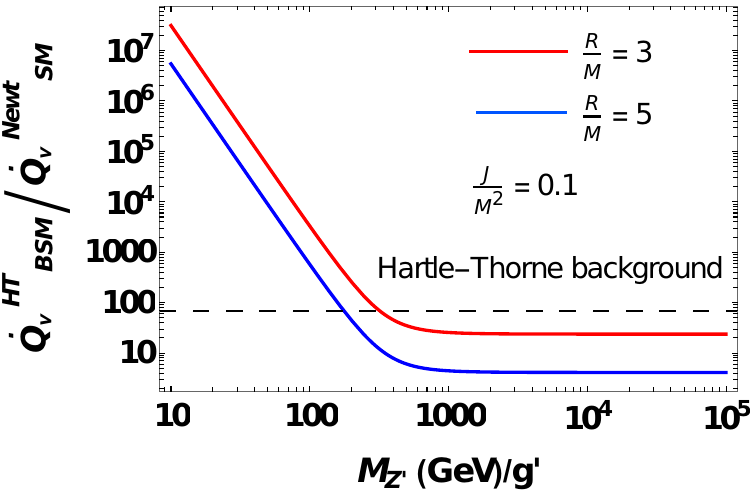}\label{plot6b}}
\caption{(a)Variation of the ratio of energy depositions in merging neutron stars in Hartle-Thorne and Newtonian background with respect to $\frac{M_{Z^\prime}}{g^\prime}$ for fixed $\frac{R}{M}$. (b)Variation of the ratio of energy depositions in merging neutron stars in Hartle-Thorne and Newtonian background with respect to $\frac{M_{Z^\prime}}{g^\prime}$ for fixed $\frac{J}{M^2}$.}
\label{plot6}
\end{figure}

In Fig.\ref{plot6} we have shown the variation of the ratio of energy depositions in merging neutron stars in Hartle-Thorne and Newtonian background with respect to $\frac{M_{Z^\prime}}{g^\prime}$ for fixed $\frac{R}{M}=3$ (Fig.\ref{plot6a}) and fixed $\frac{J}{M^2}=0.1$ (Fig.\ref{plot6b}). In both cases, the energy deposition ratio increases with decreasing $\frac{R}{M}$ and $\frac{J}{M^2}$. 
In Fig.\ref{plot6a}, the red and blue lines denote the variation of the energy deposition ratio for the values $\frac{J}{M^2}=0.1$ and $0.8$ respectively for fixed $\frac{R}{M}=3$. The ratio becomes constant for $\frac{M_{Z^\prime}}{g^\prime}\gtrsim 10^3$ which corresponds to the energy enhancement factors $\sim 24$ for $\frac{J}{M^2}=0.1$ and $\sim 12$ for $\frac{J}{M^2}=0.8$. Similarly, in Fig.\ref{plot6b}, the red and blue lines denote the variation of the energy deposition ratio for the values $\frac{R}{M}=3$ and $5$ respectively for fixed $\frac{J}{M^2}=0.1$. The ratio becomes constant for $\frac{M_{Z^\prime}}{g^\prime}\gtrsim 10^3$ which corresponds to the energy enhancement factors $\sim 24$ for $\frac{R}{M}=3$ and $\sim 4$ for $\frac{R}{M}=5$.
One can enhance the GRB energy by decreasing $\frac{R}{M}$ and $\frac{J}{M^2}$. The black dashed line denotes the enhancement in energy required with respect to that of the SM+Newtonian background to explain the maximum energy in GRB.
\begin{figure}[!htbp]
\centering
\subfigure[${{\dot{Q}}^{\rm{HT}}}_{\rm{BSM}}$ vs. $\frac{R}{M}$, $\frac{J}{M^2}=0.1$]{\includegraphics[width=8cm]{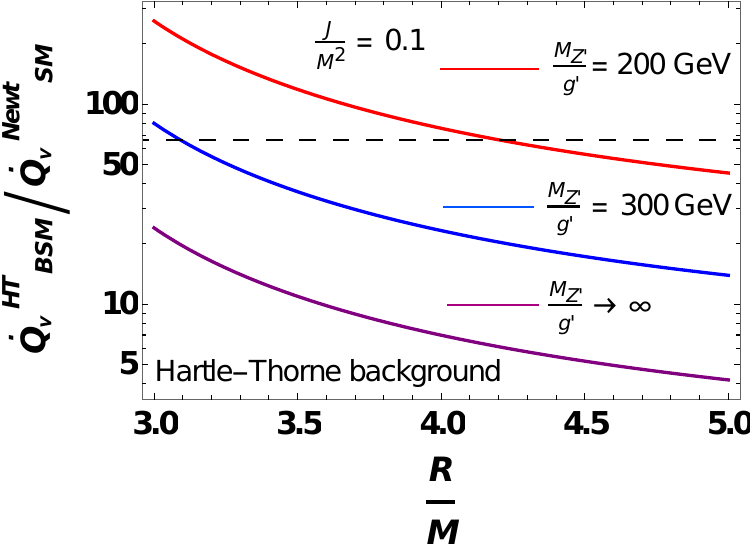}\label{plot7a}}
\subfigure[${{\dot{Q}}^{\rm{HT}}}_{\rm{BSM}}$ vs. $\frac{R}{M}$, $\frac{J}{M^2}=0.8$]{\includegraphics[width=8cm]{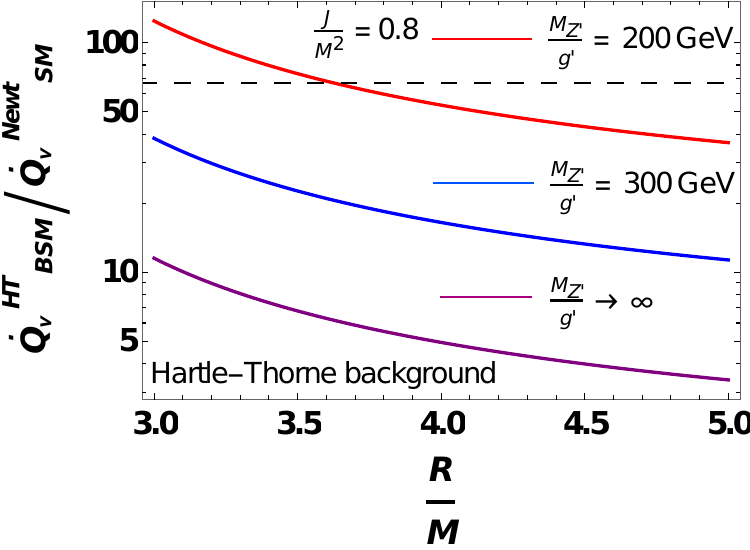}\label{plot7b}}
\caption{(a)Variation of the ratio of energy depositions in merging neutron stars in Hartle-Thorne and Newtonian background with respect to $\frac{R}{M}$ for $\frac{J}{M^2}=0.1$. (b)Variation of the ratio of energy depositions in merging neutron stars in Hartle-Thorne and Newtonian background with respect to $\frac{R}{M}$ for $\frac{J}{M^2}=0.8$.}
\label{plot7}
\end{figure}

In Fig.\ref{plot7} we have shown the variation of the ratio of energy depositions in merging neutron stars in Hartle-Thorne and Newtonian background with respect to $\frac{R}{M}$ for two fixed values of $\frac{J}{M^2}=0.1$ (Fig.\ref{plot7a}), $\frac{J}{M^2}=0.8$ (Fig.\ref{plot7b}) and with different values of $\frac{M_{Z^\prime}}{g^\prime}$. The red, blue and purple lines denote the variations with $\frac{R}{M}$ for $\frac{M_{Z^\prime}}{g^\prime}=200\hspace{0.1cm}\rm{GeV}, 300\hspace{0.1cm}\rm{GeV},$ and $ \infty$ respectively. The energy deposition ratio increases with decreasing $\frac{M_{Z^\prime}}{g^\prime}$. The value $\frac{M_{Z^\prime}}{g\prime}\rightarrow \infty$ corresponds to the SM case. If there is no BSM effect, then the enhancement in energy deposition is due to only the HT geometry that is a factor $\sim 24$ for $\frac{R}{M}=3, \frac{J}{M^2}=0.1$ (Fig.\ref{plot7a}) and a factor $\sim12$ for $\frac{R}{M}=3, \frac{J}{M^2}=0.8$ (Fig.\ref{plot7b}). The black dashed line denotes the enhancement in energy required with respect to that of the SM+Newtonian background to explain the maximum energy in GRB.  

\subsection{Born-Infeld generalization of Reissner-Nordstrom solution}
\begin{figure}[!htbp]
\centering
\subfigure[${{\dot{Q}}^{\rm{BIRN}}}_{\rm{BSM}}$ vs. $\frac{M_{Z^\prime}}{g^\prime}$, fixed $Q, b$]{\includegraphics[width=8cm]{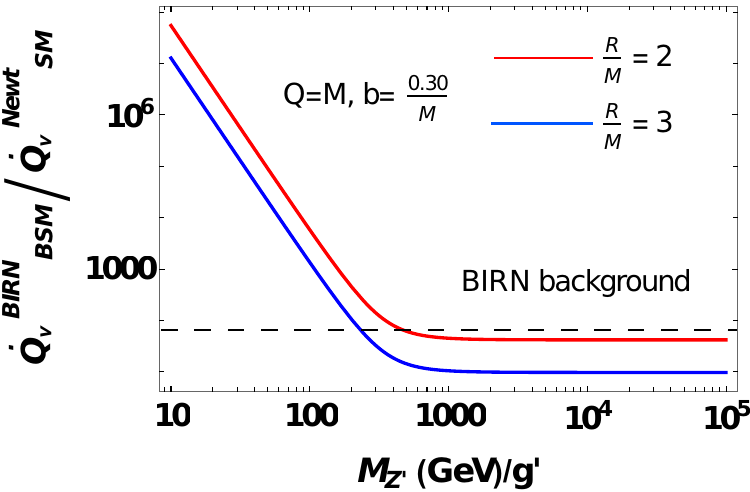}\label{birna}}
\subfigure[${{\dot{Q}}^{\rm{BIRN}}}_{\rm{BSM}}$ vs. $\frac{M_{Z^\prime}}{g^\prime}$, fixed $Q, \frac{R}{M}$]{\includegraphics[width=8cm]{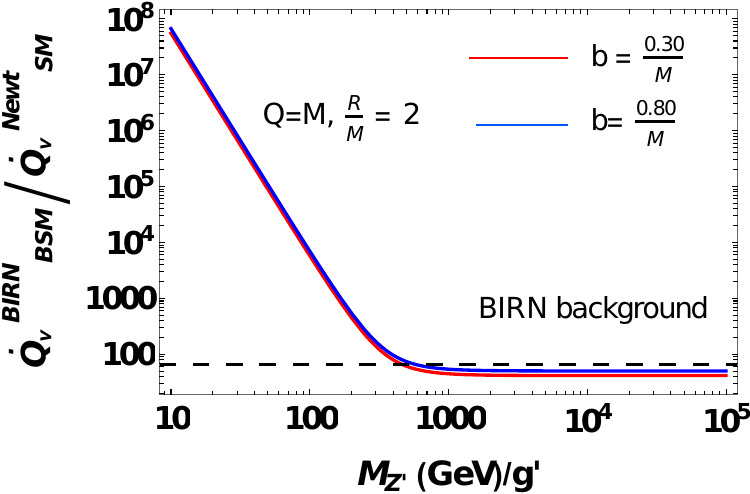}\label{birnb}}
\caption{(a)Variation of the ratio of energy depositions in Born-Infeld Reissner-Nordstrom background with respect to $\frac{M_{Z^\prime}}{g^\prime}$ for fixed $Q, b$. (b)Variation of the ratio of energy depositions Born-Infeld Reissner-Nordstrom background with respect to $\frac{M_{Z^\prime}}{g^\prime}$ for fixed $Q, \frac{R}{M}$.}
\label{1birn}
\end{figure}
In Fig.\ref{1birn} we obtain the variation of the ratio of energy deposition rate in BIRN and Newtonian background with respect to $\frac{M_{Z^\prime}}{g^\prime}$ for fixed values of $Q, b$ (Fig.\ref{birna}) and $Q, \frac{R}{M}$ (Fig\ref{birnb}). The ratio increases with increasing the values of $b$. The ratio increases as well with decreasing the values of $\frac{R}{M}$ and $\frac{M_{Z^\prime}}{g^\prime}$. The black dashed line denotes the enhancement in energy required with respect to that of the SM+Newtonian background to explain the maximum energy in GRB.
\begin{figure}[!htbp]
\centering
\subfigure[${{\dot{Q}}^{\rm{BIRN}}}_{\rm{BSM}}$ vs. $\frac{M_{Z^\prime}}{g^\prime}$, fixed $Q, b$]{\includegraphics[width=8cm]{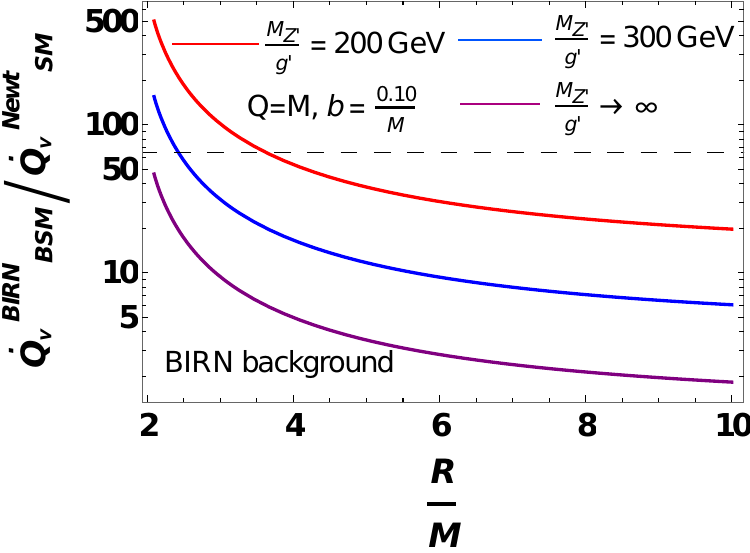}\label{2birna}}
\subfigure[${{\dot{Q}}^{\rm{BIRN}}}_{\rm{BSM}}$ vs. $\frac{M_{Z^\prime}}{g^\prime}$, fixed $Q, \frac{R}{M}$]{\includegraphics[width=8cm]{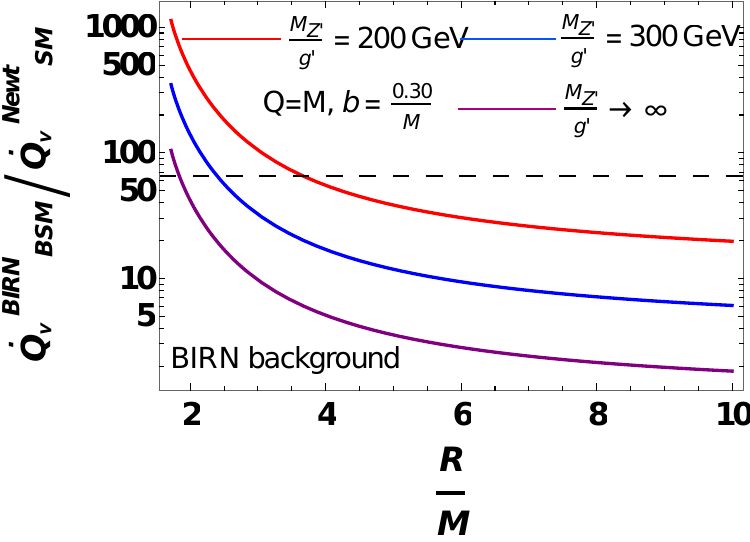}\label{2birnb}}
\caption{(a)Variation of the ratio of energy depositions in Born-Infeld Reissner-Nordstrom background with respect to $\frac{R}{M}$ for fixed $Q=M, b=\frac{0.10}{M}$. (b)Variation of the ratio of energy depositions Born-Infeld Reissner-Nordstrom background with respect to $\frac{R}{M}$ for fixed $Q=M, b=\frac{0.30}{M}$.}
\label{2birn}
\end{figure}

In Fig.\ref{2birn} we obtain the variation of the ratio of energy depositions in Born-Infeld Reissner-Nordstrom background with respect to $\frac{R}{M}$ for $Q=M, b=\frac{0.10}{M}$ (Fig.\ref{2birna}) and for $Q=M, b=\frac{0.30}{M}$ (Fig.\ref{2birnb}). The energy deposition rate increases with increasing $b$. The rate also increases with decreasing $\frac{R}{M}$ and $\frac{M_{Z^\prime}}{g^\prime}$. The black dashed line denotes the enhancement in energy required with respect to that of the SM+Newtonian background to explain the maximum energy in GRB.
\subsection{Quintessence background}
\begin{figure}[!htbp]
\centering
\subfigure[${{\dot{Q}}^{\rm{Quint}}}_{\rm{BSM}}$ vs. $\frac{M_{Z^\prime}}{g^\prime}$, fixed $\frac{R}{M}, \omega$]{\includegraphics[width=8cm]{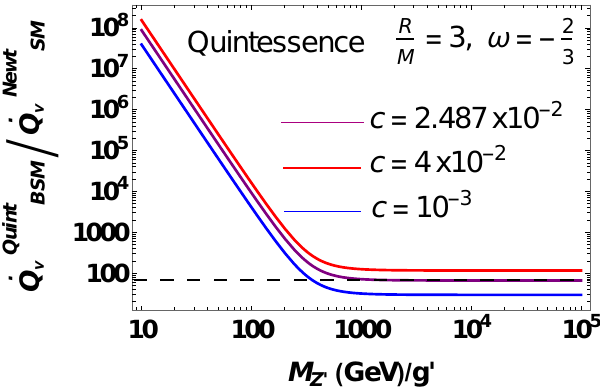}\label{w1quint}}
\subfigure[${{\dot{Q}}^{\rm{Quint}}}_{\rm{BSM}}$ vs. $\frac{M_{Z^\prime}}{g^\prime}$, fixed $\frac{R}{M}, \omega$]{\includegraphics[width=8cm]{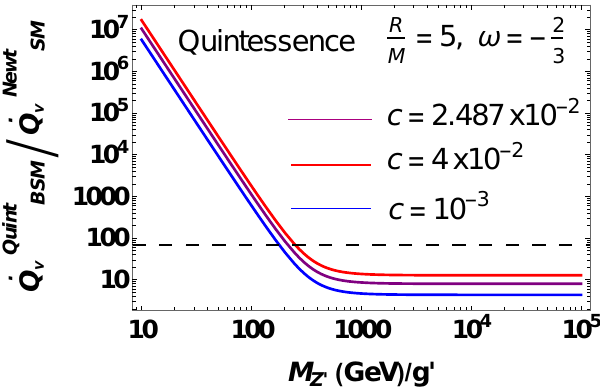}\label{w2quint}}
\subfigure[${{\dot{Q}}^{\rm{Quint}}}_{\rm{BSM}}$ vs. $\frac{M_{Z^\prime}}{g^\prime}$, fixed $\frac{R}{M}, \omega$]{\includegraphics[width=8cm]{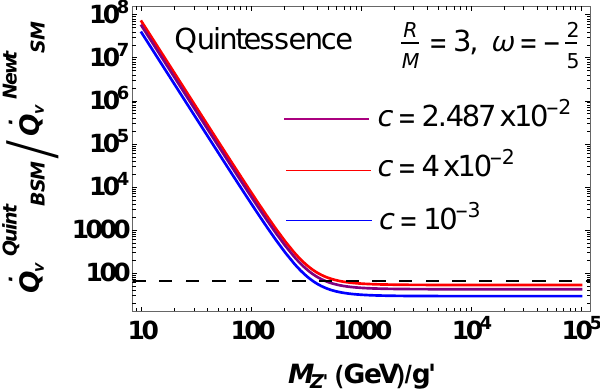}\label{w3quint}}
\caption{Variation of the ratio of energy depositions in Quintessence surrounded black hole solution with respect to $\frac{M_{Z^\prime}}{g^\prime}$ for different values of quintessence parameter with (a) $\frac{R}{M}=3, \omega=-\frac{2}{3}$, (b) $\frac{R}{M}=5, \omega=-\frac{2}{3}$ and (c) $\frac{R}{M}=3, \omega=-\frac{2}{5}$.}
\label{wquint}
\end{figure}
In Fig.\ref{wquint} we obtain the variation of the ratio of energy deposition rate in Quintessence and Newtonian background with respect to $\frac{M_{Z^\prime}}{g^\prime}$ for different values of quintessence parameters. We have shown the variation for $\frac{R}{M}=3, \omega=-\frac{2}{3}$ (Fig.\ref{w1quint}), $\frac{R}{M}=5, \omega=-\frac{2}{3}$ (Fig.\ref{w2quint}) and $\frac{R}{M}=3, \omega=-\frac{2}{5}$ (Fig.\ref{w3quint}). The energy deposition rate increases with increasing the value of quintessence parameter $c$. Also, the rate decreases with increasing the value of $\frac{R}{M}$ and $\omega$. 
\begin{figure}[!htbp]
\centering
\subfigure[${{\dot{Q}}^{\rm{Quint}}}_{\rm{BSM}}$ vs. $\frac{R}{M}$, fixed $c, \omega$]{\includegraphics[width=8cm]{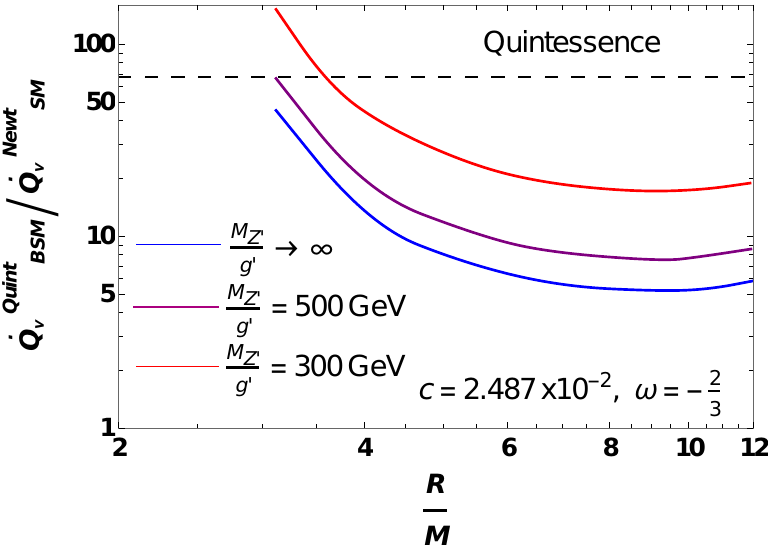}\label{w4quint}}
\subfigure[${{\dot{Q}}^{\rm{Quint}}}_{\rm{BSM}}$ vs. $\frac{R}{M}$, fixed $c, \omega$]{\includegraphics[width=8cm]{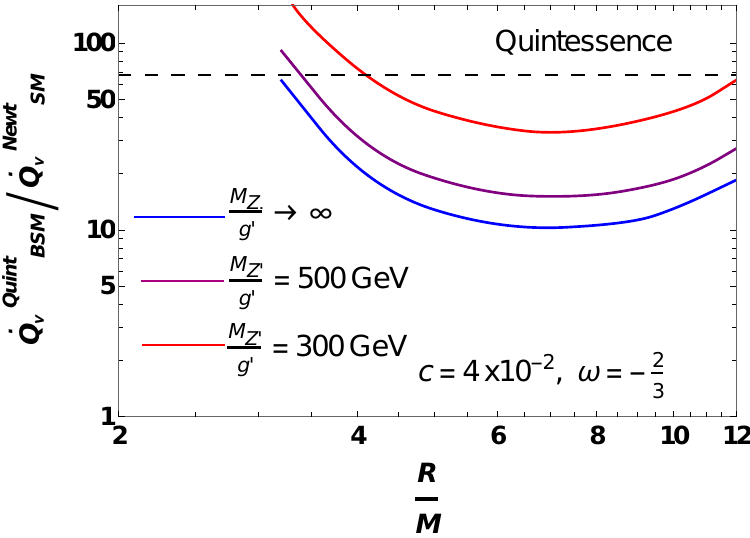}\label{w5quint}}
\subfigure[${{\dot{Q}}^{\rm{Quint}}}_{\rm{BSM}}$ vs. $\frac{R}{M}$, fixed $c, \omega$]{\includegraphics[width=8cm]{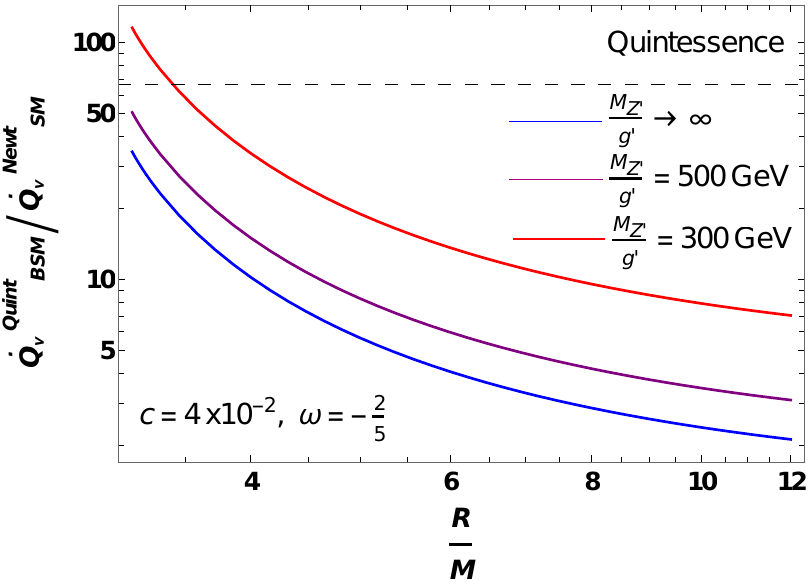}\label{w6quint}}
\caption{Variation of the ratio of energy depositions in Quintessence field surrounded black hole solution with respect to $\frac{R}{M}$ for different values of $\frac{M_{Z^\prime}}{g^\prime}$ with (a) $c=2.487\times 10^{-2}, \omega=-\frac{2}{3}$, (b) $c=4\times 10^{-2}, \omega=-\frac{2}{3}$ and (c) $c=4\times 10^{-2}, \omega=-\frac{2}{5}$.}
\label{xquint}
\end{figure}
In Fig.\ref{xquint} we obtain the variation of the ratio of energy deposition rate in Quintessence and Newtonian background with respect to $\frac{R}{M}$ for different values of $\frac{M_{Z^\prime}}{g^\prime}$. We have shown the variation for $c=2.487\times 10^{-2}, \omega=-\frac{2}{3}$ (Fig.\ref{w4quint}), $c=4\times 10^{-2}, \omega=-\frac{2}{3}$ (Fig.\ref{w5quint}), and $c=4\times 10^{-2}, \omega=-\frac{2}{5}$ (Fig.\ref{w6quint}). The blue line in Fig.\ref{xquint} corresponds to the SM scenario which means the energy deposition is only enhanced by the quintessence background. The energy deposition decreases with increasing the value of $\frac{M_{Z^\prime}}{g^\prime}$.
\section{Constraints on $Z^\prime$ from GRB observations in Newtonian, Schwarzschild, Hartle-Thorne, Born-Infeld generalization of Reissner-Nordstrom solution, and Quintessence backgrounds }\label{combined}
\begin{figure}[h]
\includegraphics[height=8cm]{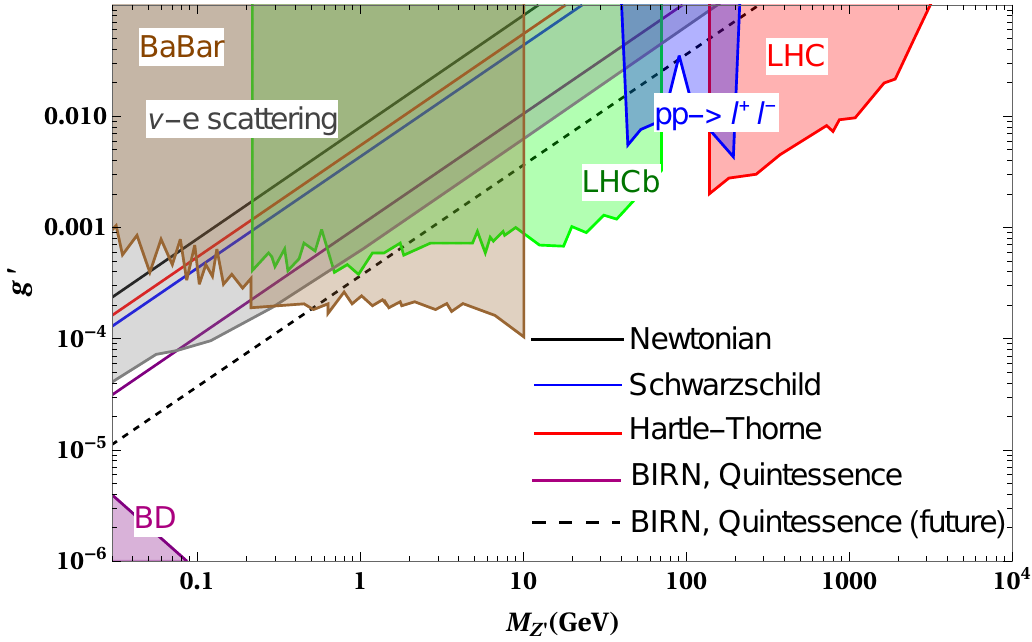}
\caption{Parameter space in $g^\prime-M_{Z^\prime}$ plane. The existing constraints are from BaBar (brown), LHCb (green), LHC (red), Beam-Dump (purple), Drell-Yan process from ATLAS (blue), and $\nu-e$ scattering (grey). The regions above the black, red, blue, and purple lines corresponding to Newtonian, Hartle-Thorne, Schwarzschild, and BIRN/Quintessence backgrounds respectively are excluded.}
\label{global}
\end{figure}
In this section, we obtain constraints on the $Z^\prime$ gauge boson from the observation of GRB in Newtonian, Schwarzschild, Hartle-Thorne, Born-Infeld Reissner-Nordstrom gravity, and Quintessence backgrounds. This is presented in Fig.\ref{global}. We also compare these bounds with the existing constraints from BaBar \cite{PhysRevLett.113.201801} (brown), LHCb \cite{PhysRevLett.120.061801} (green), LHC \cite{ATLAS:2017fih} (red), Beam-Dump experiments (purple) \cite{Ilten:2018crw,Bjorken:2009mm,Andreas:2012mt,Blumlein:2013cua} , Drell-Yan process from ATLAS \cite{ATLAS:2017rue,Escudero:2018fwn} (blue), and $\nu-e$ scattering \cite{PhysRevD.92.033009,Lindner2018} (grey). The bounds on gauge coupling $g^\prime$ for different backgrounds are obtained considering the maximum GRB energy as $10^{52}~\rm{erg}$. The black, blue, red, and purple solid lines denote the variation of $g^\prime$ with $M_{Z^\prime}$ for Newtonian, Schwarzschild, Hartle-Thorne, Born-Infeld Reissner-Nordstrom gravity, and Quintessence backgrounds respectively. The regions above these lines correspond to the parameter space for which the energy deposition due to neutrino heating is $>10^{52}~\rm{erg}$ and are excluded. The contribution of $Z^\prime$ in powering a GRB should be within the difference of energy depositions as measured from observation and from theoretically calculated SM mediated process, i.e; $\dot{Q}^{\rm{BSM}}_{\rm{b}}\lesssim\dot{Q}_{\rm{obs}}-\dot{Q}^{\rm{SM}}_{\rm{b}}$, where $b$ denotes the different backgrounds. For Newtonian, Schwarzschild, and Hartle-Thorne backgrounds, the value of $\dot{Q}^{\rm{SM}}_b\ll\dot{Q}^{\rm{obs}}$. On the other hand, for BIRN and Quintessence backgrounds, $\dot{Q}_{\rm{obs}}\sim\dot{Q}^{\rm{SM}}_{\rm{b}}$ and hence we get $\dot{Q}^{\rm{BSM}}_{\rm{b}}\lesssim \sigma$, where $\sigma$ is the observational uncertainty.

We obtain the bounds on $g^\prime$ for certain choices of model parameters. For, Schwarzschild background, we have taken $\frac{R}{M}=3$ and for Hartle-Thorne background, we have chosen $\frac{R}{M}=3$ and $\frac{J}{M^2}=0.1$. The bound on gauge coupling gets weaker as one increases the values of $\frac{R}{M}$ and $\frac{J}{M^2}$.

We obtain bounds on $g^\prime$ considering $\frac{R}{M}=2, Q=M, b=\frac{0.30}{M}$ for BIRN background whereas for Quintessence, we have chosen $\frac{R}{M}=3, \omega=-\frac{2}{3}, c=2.487\times 10^{-2}$.  One can get a weaker bound on $g^\prime$ in the BIRN background by decreasing the Born-Infeld parameter. In the Quintessence background, the bound becomes weaker as one increases the value of $\omega$ and decreases the quintessence parameter $c$ . The contribution of $Z^\prime$ in the Born-Infeld Reissner-Nordstrom solution and Quintessence background for neutrino heating is limited to be no larger than the uncertainty $(<10\%)$ in the GRB energy measurement.

We obtain the weakest bound on $g^\prime$ for Newtonian background, and the strongest bound for BIRN and Quintessence background from GRB observation. The bounds on $g^\prime$ in Newtonian, Schwarzschild, and Hartle-Thorne backgrounds are weaker compared to the neutrino-electron scattering and collider experiments. For the Born-Infeld Reissner-Nordstrom gravity and the Quintessence background, the bounds on $g^\prime$ are stronger than neutrino-electron scattering experiment for $0.03~\rm{GeV}\lesssim M_{Z^\prime}\lesssim 0.07~\rm{GeV}$. Here, we have chosen the conservative limit $M_{Z^\prime}\gtrsim 0.03~\rm{GeV}$ as the mass of the gauge boson should be greater than the centre of mass energy $(\sqrt{s}\approx 2E_\nu\approx 0.02~\rm{GeV})$ of neutrino-antineutrino pair annihilation. Future observation of GRB with better accuracy $(<1\%)$ will yield stronger bounds on the gauge coupling compared to the $\nu-e$ scattering experiments for the mass range of gauge boson $0.03~\rm{GeV}\lesssim M_{Z^\prime}\lesssim 0.5~\rm{GeV}$ as shown by the black dashed line for BIRN and quintessence backgrounds. For $M_{Z^\prime}\gtrsim 0.5~\rm{GeV}$, collider experiments give stronger bounds on gauge coupling.
\section{conclusion}\label{sec6}
In this paper, we obtain constraints on gauge coupling of $Z^\prime$ gauge boson from the GRB observation considering neutrino pair annihilation is the only process powering GRB. We derive the constraints on gauge coupling for Newtonian, Schwarzschild, Hartle-Thorne, Born-Infeld Reissner Nordstrom gravity, and Quintessence background. The neutrino pair annihilation in the Newtonian background cannot deposit enough amount of energy $(\sim 1.5\times 10^{50}~\rm{erg})$ to attain its maximum value $(\sim 10^{52}~\rm{erg})$ as measured from the GRB observations. In Schwarzschild and Hartle-Thorne backgrounds, the energy deposition due to neutrino heating increases by factors $\sim 30$ and $\sim 24$ as compared to the Newtonian background, still the maximum observed GRB energy cannot be reached. In modified gravity theories such as the Born-Infeld generalization of Reissner-Nordstrom solution and Quintessence, one can attain the maximum GRB energy for certain choices of modified gravity parameters.  

Here, we consider the $U(1)_{\rm{B-L}}$ extension of the SM where the $Z^\prime$ gauge boson of the $U(1)_{\rm{B-L}}$ gauge group can also mediate neutrino pair annihilation process. We calculate the energy deposition due to $Z^\prime$ mediated neutrino heating in all the above mentioned backgrounds. Compared with the GRB observational data, we obtain constraints on gauge coupling of $Z^\prime$ in all the backgrounds. The bounds in Newtonian, Schwarzschild, and Hartle-Thorne backgrounds are weaker than the neutrino-electron scattering and collider experiments whereas for a certain range of $M_{Z^\prime}$ we obtain stronger bound on gauge coupling for Born-Infeld generalization of Reissner-Nordstrom solution and quintessence backgrounds. Future observations with better accuracy can improve our results for modified gravity backgrounds up to one order of magnitude. Moreover, it is important to mention here that we have not taken the effects of any trapping of neutrinos \cite{Ghosh:1995dn} and nonlinear magnetic fields \cite{Jamil:2014rsa,MosqueraCuesta:2017iln} that can significantly change the energy deposition rate. In conclusion, our study highlights the importance of the inclusion of BSM contribution to the neutrino annihilation process in constraining new physics. This can open up new avenues in understanding the physics of GRBs.
\section*{Acknowledgements}
S.G. acknowledges the J.C Bose Fellowship (JCB/2020/000011) of Science and Engineering Research Board of Department of Science and Technology, Government of India.
AKM acknowledges the support through the Ramanujan Fellowship (PI: Dr. Diptimoy Ghosh) offered by the Department of Science and Technology, Government of India. AKM would also like to acknowledge the hospitality at Physical Research Laboratory, Ahmedabad, and support via J.C Bose Fellowship (JCB/2020/000011) during the initial stage of the work.
\bibliographystyle{utphys}
\bibliography{lumi}

\providecommand{\href}[2]{#2}\begingroup\raggedright\begin{thebibliography}{10}

\bibitem{Bethe:1990mw}
H.~A. Bethe, ``{Supernova mechanisms},''
  \href{http://dx.doi.org/10.1103/RevModPhys.62.801}{{\em Rev. Mod. Phys.}
  {\bfseries 62} (1990) 801--866}.

\bibitem{Goodman:1986we}
J.~Goodman, A.~Dar, and S.~Nussinov, ``{NEUTRINO ANNIHILATION IN TYPE II
  SUPERNOVAE},'' \href{http://dx.doi.org/10.1086/184840}{{\em Astrophys. J.
  Lett.} {\bfseries 314} (1987) L7--L10}.

\bibitem{Bethe:1985sox}
H.~A. Bethe and J.~R. Wilson, ``{Revival of a stalled supernova shock by
  neutrino heating},'' \href{http://dx.doi.org/10.1086/163343}{{\em Astrophys.
  J.} {\bfseries 295} (1985) 14--23}.

\bibitem{Bhattacharya1991}
D.~Bhattacharya and E.~P.~J. van~den Heuvel, ``Formation and evolution of
  binary and millisecond radio pulsars,''
  \href{http://dx.doi.org/10.1016/0370-1573(91)90064-S}{{\em Physics Reports}
  {\bfseries 203} (Jan, 1991) 1--124}.
  \url{https://ui.adsabs.harvard.edu/abs/1991PhR...203....1B}.

\bibitem{Mathews:1997vw}
G.~J. Mathews and J.~R. Wilson, ``{Binary induced neutron star compression,
  heating, and collapse},'' \href{http://dx.doi.org/10.1086/304166}{{\em
  Astrophys. J.} {\bfseries 482} (1997) 929--941},
  \href{http://arxiv.org/abs/astro-ph/9701142}{{\ttfamily
  arXiv:astro-ph/9701142}}.

\bibitem{Salmonson:1999es}
J.~D. Salmonson and J.~R. Wilson, ``{General relativistic augmentation of
  neutrino pair annihilation energy deposition near neutron stars},''
  \href{http://dx.doi.org/10.1086/307232}{{\em Astrophys. J.} {\bfseries 517}
  (1999) 859--865}, \href{http://arxiv.org/abs/astro-ph/9908017}{{\ttfamily
  arXiv:astro-ph/9908017}}.

\bibitem{Salmonson:2001tz}
J.~D. Salmonson and J.~R. Wilson, ``{Neutrino annihilation between binary
  neutron stars},'' \href{http://dx.doi.org/10.1086/323319}{{\em Astrophys. J.}
  {\bfseries 561} (2001) 950--956},
  \href{http://arxiv.org/abs/astro-ph/0108196}{{\ttfamily
  arXiv:astro-ph/0108196}}.

\bibitem{Asano:2000ib}
K.~Asano and T.~Fukuyama, ``{Neutrino pair annihilation in the gravitation of
  gamma-ray burst sources},'' \href{http://dx.doi.org/10.1086/308513}{{\em
  Astrophys. J.} {\bfseries 531} (2000) 949--955},
  \href{http://arxiv.org/abs/astro-ph/0002196}{{\ttfamily
  arXiv:astro-ph/0002196}}.

\bibitem{Asano:2000dq}
K.~Asano and T.~Fukuyama, ``{Relativistic effects on neutrino pair annihilation
  above a Kerr black hole with the accretion disk},''
  \href{http://dx.doi.org/10.1086/318312}{{\em Astrophys. J.} {\bfseries 546}
  (2001) 1019--1026}, \href{http://arxiv.org/abs/astro-ph/0009453}{{\ttfamily
  arXiv:astro-ph/0009453}}.

\bibitem{Prasanna:2001ie}
A.~R. Prasanna and S.~Goswami, ``{Energy deposition due to neutrino pair
  annihilation near rotating neutron stars},''
  \href{http://dx.doi.org/10.1016/S0370-2693(01)01470-8}{{\em Phys. Lett. B}
  {\bfseries 526} (2002) 27--33},
  \href{http://arxiv.org/abs/astro-ph/0109058}{{\ttfamily
  arXiv:astro-ph/0109058}}.

\bibitem{Birkl:2006mu}
R.~Birkl, M.~A. Aloy, H.~T. Janka, and E.~Mueller, ``{Neutrino pair
  annihilation near accreting, stellar-mass black holes},''
  \href{http://dx.doi.org/10.1051/0004-6361:20066293}{{\em Astron. Astrophys.}
  {\bfseries 463} (2007) 51},
  \href{http://arxiv.org/abs/astro-ph/0608543}{{\ttfamily
  arXiv:astro-ph/0608543}}.

\bibitem{Miller:2002hqu}
W.~A. Miller, N.~D. George, A.~Kheyfets, and J.~M. McGhee, ``{Off-axis neutrino
  scattering in grb central engines},''
  \href{http://dx.doi.org/10.1086/345471}{{\em Astrophys. J.} {\bfseries 583}
  (2003) 833--841}, \href{http://arxiv.org/abs/astro-ph/0205213}{{\ttfamily
  arXiv:astro-ph/0205213}}.

\bibitem{Meszaros:1992gc}
P.~Meszaros and M.~J. Rees, ``{High entropy fireballs and jets in gamma-ray
  burst sources},'' {\em Mon. Not. Roy. Astron. Soc.} {\bfseries 257} (1992)
  29--31.

\bibitem{Ruffert:1998qg}
M.~Ruffert and H.~T. Janka, ``{Gamma-ray bursts from accreting black holes in
  neutron star mergers},'' {\em Astron. Astrophys.} {\bfseries 344} (1999)
  573--606, \href{http://arxiv.org/abs/astro-ph/9809280}{{\ttfamily
  arXiv:astro-ph/9809280}}.

\bibitem{Zhang:2009ew}
D.~Zhang and Z.~G. Dai, ``{Hyperaccreting Disks around Magnetars for Gamma-Ray
  Bursts: Effects of Strong Magnetic Fields},''
  \href{http://dx.doi.org/10.1088/0004-637X/718/2/841}{{\em Astrophys. J.}
  {\bfseries 718} (2010) 841--866},
  \href{http://arxiv.org/abs/0911.5528}{{\ttfamily arXiv:0911.5528
  [astro-ph.HE]}}.

\bibitem{Kovacs:2010zp}
Z.~Kovacs, K.~S. Cheng, and T.~Harko, ``{Electron-positron energy deposition
  rate from neutrino pair annihilation on the rotation axis of neutron and
  quark stars},''
  \href{http://dx.doi.org/10.1111/j.1365-2966.2010.17784.x}{{\em Mon. Not. Roy.
  Astron. Soc.} {\bfseries 411} (2011) 1503--1524},
  \href{http://arxiv.org/abs/1009.6029}{{\ttfamily arXiv:1009.6029
  [astro-ph.HE]}}.

\bibitem{Lambiase:2020iul}
G.~Lambiase and L.~Mastrototaro, ``{Effects of modified theories of gravity on
  neutrino pair annihilation energy deposition near neutron stars},''
  \href{http://dx.doi.org/10.3847/1538-4357/abba2c}{{\em Astrophys. J.}
  {\bfseries 904} no.~1, (2020) 19},
  \href{http://arxiv.org/abs/2009.08722}{{\ttfamily arXiv:2009.08722
  [astro-ph.HE]}}.

\bibitem{Lambiase:2020pkc}
G.~Lambiase and L.~Mastrototaro, ``{Neutrino pair annihilation ($\nu {\bar{\nu
  }}\rightarrow e^-e^+$) in the presence of quintessence surrounding a black
  hole},'' \href{http://dx.doi.org/10.1140/epjc/s10052-021-09732-2}{{\em Eur.
  Phys. J. C} {\bfseries 81} no.~10, (2021) 932},
  \href{http://arxiv.org/abs/2012.09100}{{\ttfamily arXiv:2012.09100
  [astro-ph.HE]}}.

\bibitem{Paczynski1990}
B.~Paczynski, ``Super-eddington winds from neutron stars,''
  \href{http://dx.doi.org/10.1086/169332}{{\em The Astrophysical Journal}
  {\bfseries 363} (Nov, 1990) 218}.
  \url{https://ui.adsabs.harvard.edu/abs/1990ApJ...363..218P}.

\bibitem{Cooperstein1987}
J.~Cooperstein, L.~J. van~den Horn, and E.~Baron, ``Neutrino pair energy
  deposition in supernovae,'' \href{http://dx.doi.org/10.1086/185019}{{\em The
  Astrophysical Journal} {\bfseries 321} (Oct, 1987) L129}.
  \url{https://ui.adsabs.harvard.edu/abs/1987ApJ...321L.129C}.

\bibitem{Eichler:1989ve}
D.~Eichler, M.~Livio, T.~Piran, and D.~N. Schramm, ``{Nucleosynthesis, Neutrino
  Bursts and Gamma-Rays from Coalescing Neutron Stars},''
  \href{http://dx.doi.org/10.1038/340126a0}{{\em Nature} {\bfseries 340} (1989)
  126--128}.

\bibitem{Leng:2014dfa}
M.~Leng and D.~Giannios, ``{Testing the neutrino annihilation model for
  launching GRB jets},'' \href{http://dx.doi.org/10.1093/mnrasl/slu122}{{\em
  Mon. Not. Roy. Astron. Soc.} {\bfseries 445} (2014) 1},
  \href{http://arxiv.org/abs/1408.4509}{{\ttfamily arXiv:1408.4509
  [astro-ph.HE]}}.

\bibitem{Zhang:2003uk}
B.~Zhang and P.~Meszaros, ``{Gamma-ray bursts: Progress, problems \&
  prospects},'' \href{http://dx.doi.org/10.1142/S0217751X0401746X}{{\em Int. J.
  Mod. Phys. A} {\bfseries 19} (2004) 2385--2472},
  \href{http://arxiv.org/abs/astro-ph/0311321}{{\ttfamily
  arXiv:astro-ph/0311321}}.

\bibitem{Piran:2004ba}
T.~Piran, ``{The physics of gamma-ray bursts},''
  \href{http://dx.doi.org/10.1103/RevModPhys.76.1143}{{\em Rev. Mod. Phys.}
  {\bfseries 76} (2004) 1143--1210},
  \href{http://arxiv.org/abs/astro-ph/0405503}{{\ttfamily
  arXiv:astro-ph/0405503}}.

\bibitem{Luongo:2021pjs}
O.~Luongo and M.~Muccino, ``{A Roadmap to Gamma-Ray Bursts: New Developments
  and Applications to Cosmology},''
  \href{http://dx.doi.org/10.3390/galaxies9040077}{{\em Galaxies} {\bfseries 9}
  no.~4, (2021) 77}, \href{http://arxiv.org/abs/2110.14408}{{\ttfamily
  arXiv:2110.14408 [astro-ph.HE]}}.

\bibitem{Perego:2017fho}
A.~Perego, H.~Yasin, and A.~Arcones, ``{Neutrino pair annihilation above merger
  remnants: implications of a long-lived massive neutron star},''
  \href{http://dx.doi.org/10.1088/1361-6471/aa7bdc}{{\em J. Phys. G} {\bfseries
  44} no.~8, (2017) 084007}, \href{http://arxiv.org/abs/1701.02017}{{\ttfamily
  arXiv:1701.02017 [astro-ph.HE]}}.

\bibitem{Breton:2002td}
N.~Breton, ``{Geodesic structure of the Born-Infeld black hole},''
  \href{http://dx.doi.org/10.1088/0264-9381/19/4/301}{{\em Class. Quant. Grav.}
  {\bfseries 19} (2002) 601--612}.

\bibitem{Babichev:2015rva}
E.~Babichev, C.~Charmousis, and M.~Hassaine, ``{Charged Galileon black
  holes},'' \href{http://dx.doi.org/10.1088/1475-7516/2015/05/031}{{\em JCAP}
  {\bfseries 05} (2015) 031}, \href{http://arxiv.org/abs/1503.02545}{{\ttfamily
  arXiv:1503.02545 [gr-qc]}}.

\bibitem{BeltranJimenez:2017doy}
J.~Beltran~Jimenez, L.~Heisenberg, G.~J. Olmo, and D.~Rubiera-Garcia,
  ``{Born\textendash{}Infeld inspired modifications of gravity},''
  \href{http://dx.doi.org/10.1016/j.physrep.2017.11.001}{{\em Phys. Rept.}
  {\bfseries 727} (2018) 1--129},
  \href{http://arxiv.org/abs/1704.03351}{{\ttfamily arXiv:1704.03351 [gr-qc]}}.

\bibitem{Sotiriou:2014pfa}
T.~P. Sotiriou and S.-Y. Zhou, ``{Black hole hair in generalized scalar-tensor
  gravity: An explicit example},''
  \href{http://dx.doi.org/10.1103/PhysRevD.90.124063}{{\em Phys. Rev. D}
  {\bfseries 90} (2014) 124063},
  \href{http://arxiv.org/abs/1408.1698}{{\ttfamily arXiv:1408.1698 [gr-qc]}}.

\bibitem{Brans:1961sx}
C.~Brans and R.~H. Dicke, ``{Mach's principle and a relativistic theory of
  gravitation},'' \href{http://dx.doi.org/10.1103/PhysRev.124.925}{{\em Phys.
  Rev.} {\bfseries 124} (1961) 925--935}.

\bibitem{Kokkotas:2017zwt}
K.~Kokkotas, R.~A. Konoplya, and A.~Zhidenko, ``{Non-Schwarzschild black-hole
  metric in four dimensional higher derivative gravity: analytical
  approximation},'' \href{http://dx.doi.org/10.1103/PhysRevD.96.064007}{{\em
  Phys. Rev. D} {\bfseries 96} no.~6, (2017) 064007},
  \href{http://arxiv.org/abs/1705.09875}{{\ttfamily arXiv:1705.09875 [gr-qc]}}.

\bibitem{Lambiase:2022ywp}
G.~Lambiase and L.~Mastrototaro, ``{Neutrino pair annihilation above black-hole
  accretion disks in modified gravity},''
  \href{http://arxiv.org/abs/2205.09785}{{\ttfamily arXiv:2205.09785
  [hep-ph]}}.

\bibitem{Shi:2022pbc}
Y.~Shi and H.~Cheng, ``{The neutrino pair annihilation
  ($\nu\bar{\nu}\longrightarrow e^{-}e^{+}$)around a massive source with an
  $f(R)$ global monopole},'' \href{http://arxiv.org/abs/2206.00670}{{\ttfamily
  arXiv:2206.00670 [gr-qc]}}.

\bibitem{Das:2021nqj}
A.~Das, S.~Goswami, V.~K.~N., and T.~K. Poddar, ``{Freeze-in sterile neutrino
  dark matter in a class of U$(1)^\prime$ models with inverse seesaw},''
  \href{http://arxiv.org/abs/2104.13986}{{\ttfamily arXiv:2104.13986
  [hep-ph]}}.

\bibitem{Chakraborty:2021apc}
K.~Chakraborty, A.~Das, S.~Goswami, and S.~Roy, ``{Constraining general U(1)
  interactions from neutrino-electron scattering measurements at DUNE near
  detector},'' \href{http://dx.doi.org/10.1007/JHEP04(2022)008}{{\em JHEP}
  {\bfseries 04} (2022) 008}, \href{http://arxiv.org/abs/2111.08767}{{\ttfamily
  arXiv:2111.08767 [hep-ph]}}.

\bibitem{Feng:2022inv}
J.~L. Feng {\em et~al.}, ``{The Forward Physics Facility at the High-Luminosity
  LHC},'' \href{http://arxiv.org/abs/2203.05090}{{\ttfamily arXiv:2203.05090
  [hep-ex]}}.

\bibitem{Erler:2009jh}
J.~Erler, P.~Langacker, S.~Munir, and E.~Rojas, ``{Improved Constraints on
  Z-prime Bosons from Electroweak Precision Data},''
  \href{http://dx.doi.org/10.1088/1126-6708/2009/08/017}{{\em JHEP} {\bfseries
  08} (2009) 017}, \href{http://arxiv.org/abs/0906.2435}{{\ttfamily
  arXiv:0906.2435 [hep-ph]}}.

\bibitem{Dittmar:2003ir}
M.~Dittmar, A.-S. Nicollerat, and A.~Djouadi, ``{Z-prime studies at the LHC: An
  Update},'' \href{http://dx.doi.org/10.1016/j.physletb.2003.09.103}{{\em Phys.
  Lett. B} {\bfseries 583} (2004) 111--120},
  \href{http://arxiv.org/abs/hep-ph/0307020}{{\ttfamily arXiv:hep-ph/0307020}}.

\bibitem{Basso:2008iv}
L.~Basso, A.~Belyaev, S.~Moretti, and C.~H. Shepherd-Themistocleous,
  ``{Phenomenology of the minimal B-L extension of the Standard model: Z' and
  neutrinos},'' \href{http://dx.doi.org/10.1103/PhysRevD.80.055030}{{\em Phys.
  Rev. D} {\bfseries 80} (2009) 055030},
  \href{http://arxiv.org/abs/0812.4313}{{\ttfamily arXiv:0812.4313 [hep-ph]}}.

\bibitem{Das:2016zue}
A.~Das, S.~Oda, N.~Okada, and D.-s. Takahashi, ``{Classically conformal U(1)'
  extended standard model, electroweak vacuum stability, and LHC Run-2
  bounds},'' \href{http://dx.doi.org/10.1103/PhysRevD.93.115038}{{\em Phys.
  Rev. D} {\bfseries 93} no.~11, (2016) 115038},
  \href{http://arxiv.org/abs/1605.01157}{{\ttfamily arXiv:1605.01157
  [hep-ph]}}.

\bibitem{Accomando:2017qcs}
E.~Accomando, L.~Delle~Rose, S.~Moretti, E.~Olaiya, and C.~H.
  Shepherd-Themistocleous, ``{Extra Higgs boson and $Z^\prime$ as portals to
  signatures of heavy neutrinos at the LHC},''
  \href{http://dx.doi.org/10.1007/JHEP02(2018)109}{{\em JHEP} {\bfseries 02}
  (2018) 109}, \href{http://arxiv.org/abs/1708.03650}{{\ttfamily
  arXiv:1708.03650 [hep-ph]}}.

\bibitem{Ekstedt:2016wyi}
A.~Ekstedt, R.~Enberg, G.~Ingelman, J.~L\"ofgren, and T.~Mandal,
  ``{Constraining minimal anomaly free $\mathrm{U}(1)$ extensions of the
  Standard Model},'' \href{http://dx.doi.org/10.1007/JHEP11(2016)071}{{\em
  JHEP} {\bfseries 11} (2016) 071},
  \href{http://arxiv.org/abs/1605.04855}{{\ttfamily arXiv:1605.04855
  [hep-ph]}}.

\bibitem{Das:2019pua}
A.~Das, S.~Goswami, K.~N. Vishnudath, and T.~Nomura, ``{Constraining a general
  U(1)$^\prime$ inverse seesaw model from vacuum stability, dark matter and
  collider},'' \href{http://dx.doi.org/10.1103/PhysRevD.101.055026}{{\em Phys.
  Rev. D} {\bfseries 101} no.~5, (2020) 055026},
  \href{http://arxiv.org/abs/1905.00201}{{\ttfamily arXiv:1905.00201
  [hep-ph]}}.

\bibitem{ATLAS:2017fih}
{\bfseries ATLAS} Collaboration, M.~Aaboud {\em et~al.}, ``{Search for new
  high-mass phenomena in the dilepton final state using $36~\rm{fb}^{−1}$ of
  proton-proton collision data at $ \sqrt{s}=13 $ TeV with the ATLAS
  detector},'' \href{http://dx.doi.org/10.1007/JHEP10(2017)182}{{\em JHEP}
  {\bfseries 10} (2017) 182}, \href{http://arxiv.org/abs/1707.02424}{{\ttfamily
  arXiv:1707.02424 [hep-ex]}}.

\bibitem{ATLAS:2017rue}
{\bfseries ATLAS} Collaboration, M.~Aaboud {\em et~al.}, ``{Measurement of the
  Drell-Yan triple-differential cross section in $pp$ collisions at $\sqrt{s} =
  8$ TeV},'' \href{http://dx.doi.org/10.1007/JHEP12(2017)059}{{\em JHEP}
  {\bfseries 12} (2017) 059}, \href{http://arxiv.org/abs/1710.05167}{{\ttfamily
  arXiv:1710.05167 [hep-ex]}}.

\bibitem{PhysRevD.92.033009}
S.~Bilmi\ifmmode~\mbox{\c{s}}\else \c{s}\fi{}, I.~Turan, T.~M. Aliev, M.~Deniz,
  L.~Singh, and H.~T. Wong, ``Constraints on dark photon from neutrino-electron
  scattering experiments,''
  \href{http://dx.doi.org/10.1103/PhysRevD.92.033009}{{\em Phys. Rev. D}
  {\bfseries 92} (Aug, 2015) 033009}.
  \url{https://link.aps.org/doi/10.1103/PhysRevD.92.033009}.

\bibitem{Lindner2018}
M.~Lindner, F.~S. Queiroz, W.~Rodejohann, and X.-J. Xu, ``Neutrino-electron
  scattering: general constraints on $z^\prime$ and dark photon models,''
  \href{http://dx.doi.org/10.1007/JHEP05(2018)098}{{\em Journal of High Energy
  Physics} {\bfseries 2018} no.~5, (May, 2018) 98}.
  \url{https://doi.org/10.1007/JHEP05(2018)098}.

\bibitem{Alekhin:2015byh}
S.~Alekhin {\em et~al.}, ``{A facility to Search for Hidden Particles at the
  CERN SPS: the SHiP physics case},''
  \href{http://dx.doi.org/10.1088/0034-4885/79/12/124201}{{\em Rept. Prog.
  Phys.} {\bfseries 79} no.~12, (2016) 124201},
  \href{http://arxiv.org/abs/1504.04855}{{\ttfamily arXiv:1504.04855
  [hep-ph]}}.

\bibitem{PhysRevD.99.095011}
{\bfseries FASER Collaboration} Collaboration, A.~Ariga, T.~Ariga, J.~Boyd,
  F.~Cadoux, D.~W. Casper, Y.~Favre, J.~L. Feng, D.~Ferrere, I.~Galon,
  S.~Gonzalez-Sevilla, S.-C. Hsu, G.~Iacobucci, E.~Kajomovitz, F.~Kling,
  S.~Kuehn, L.~Levinson, H.~Otono, B.~Petersen, O.~Sato, M.~Schott, A.~Sfyrla,
  J.~Smolinsky, A.~M. Soffa, Y.~Takubo, E.~Torrence, S.~Trojanowski, and
  G.~Zhang, ``Faser's physics reach for long-lived particles,''
  \href{http://dx.doi.org/10.1103/PhysRevD.99.095011}{{\em Phys. Rev. D}
  {\bfseries 99} (May, 2019) 095011}.
  \url{https://link.aps.org/doi/10.1103/PhysRevD.99.095011}.

\bibitem{FASER:2019aik}
{\bfseries FASER} Collaboration, A.~Ariga {\em et~al.}, ``{FASER: ForwArd
  Search ExpeRiment at the LHC},''
  \href{http://arxiv.org/abs/1901.04468}{{\ttfamily arXiv:1901.04468
  [hep-ex]}}.

\bibitem{Dent:2012mx}
J.~B. Dent, F.~Ferrer, and L.~M. Krauss, ``{Constraints on Light Hidden Sector
  Gauge Bosons from Supernova Cooling},''
  \href{http://arxiv.org/abs/1201.2683}{{\ttfamily arXiv:1201.2683
  [astro-ph.CO]}}.

\bibitem{Raffelt:2000kp}
G.~G. Raffelt, ``{Astrophysics probes of particle physics},''
  \href{http://dx.doi.org/10.1016/S0370-1573(00)00039-9}{{\em Phys. Rept.}
  {\bfseries 333} (2000) 593--618}.

\bibitem{Kazanas:2014mca}
D.~Kazanas, R.~N. Mohapatra, S.~Nussinov, V.~L. Teplitz, and Y.~Zhang,
  ``{Supernova Bounds on the Dark Photon Using its Electromagnetic Decay},''
  \href{http://dx.doi.org/10.1016/j.nuclphysb.2014.11.009}{{\em Nucl. Phys. B}
  {\bfseries 890} (2014) 17--29},
  \href{http://arxiv.org/abs/1410.0221}{{\ttfamily arXiv:1410.0221 [hep-ph]}}.

\bibitem{Balaji:2022noj}
S.~Balaji, P.~S.~B. Dev, J.~Silk, and Y.~Zhang, ``{Improved stellar limits on a
  light CP-even scalar},'' \href{http://arxiv.org/abs/2205.01669}{{\ttfamily
  arXiv:2205.01669 [hep-ph]}}.

\bibitem{Fong:2015oha}
W.-f. Fong, E.~Berger, R.~Margutti, and B.~A. Zauderer, ``{A Decade of
  Short-duration Gamma-ray Burst Broadband Afterglows: Energetics, Circumburst
  Densities, and jet Opening Angles},''
  \href{http://dx.doi.org/10.1088/0004-637X/815/2/102}{{\em Astrophys. J.}
  {\bfseries 815} no.~2, (2015) 102},
  \href{http://arxiv.org/abs/1509.02922}{{\ttfamily arXiv:1509.02922
  [astro-ph.HE]}}.

\bibitem{PhysRevD.20.776}
A.~Davidson, ``$b\ensuremath{-}l$ as the fourth color within an
  $\mathrm{SU}{(2)}_{L}\ifmmode\times\else\texttimes\fi{}\mathrm{U}{(1)}_{R}\ifmmode\times\else\texttimes\fi{}\mathrm{U}(1)$
  model,'' \href{http://dx.doi.org/10.1103/PhysRevD.20.776}{{\em Phys. Rev. D}
  {\bfseries 20} (Aug, 1979) 776--783}.
  \url{https://link.aps.org/doi/10.1103/PhysRevD.20.776}.

\bibitem{Marshak:1979fm}
R.~E. Marshak and R.~N. Mohapatra, ``{Quark - Lepton Symmetry and B-L as the
  U(1) Generator of the Electroweak Symmetry Group},''
  \href{http://dx.doi.org/10.1016/0370-2693(80)90436-0}{{\em Phys. Lett. B}
  {\bfseries 91} (1980) 222--224}.

\bibitem{Mohapatra:1980qe}
R.~N. Mohapatra and R.~E. Marshak, ``{Local B-L Symmetry of Electroweak
  Interactions, Majorana Neutrinos and Neutron Oscillations},''
  \href{http://dx.doi.org/10.1103/PhysRevLett.44.1316}{{\em Phys. Rev. Lett.}
  {\bfseries 44} (1980) 1316--1319}. [Erratum: Phys.Rev.Lett. 44, 1643 (1980)].

\bibitem{Wetterich:1981bx}
C.~Wetterich, ``{Neutrino Masses and the Scale of B-L Violation},''
  \href{http://dx.doi.org/10.1016/0550-3213(81)90279-0}{{\em Nucl. Phys. B}
  {\bfseries 187} (1981) 343--375}.

\bibitem{Masiero:1982fi}
A.~Masiero, J.~F. Nieves, and T.~Yanagida, ``{$B^-$l Violating Proton Decay and
  Late Cosmological Baryon Production},''
  \href{http://dx.doi.org/10.1016/0370-2693(82)90024-7}{{\em Phys. Lett. B}
  {\bfseries 116} (1982) 11--15}.

\bibitem{ParticleDataGroup:2020ssz}
{\bfseries Particle Data Group} Collaboration, P.~A. Zyla {\em et~al.},
  ``{Review of Particle Physics},''
  \href{http://dx.doi.org/10.1093/ptep/ptaa104}{{\em PTEP} {\bfseries 2020}
  no.~8, (2020) 083C01}.

\bibitem{Cooperstein1986}
J.~Cooperstein, L.~J. van~den Horn, and E.~A. Baron, ``Neutrino flows in
  collapsing stars: A two-fluid model,''
  \href{http://dx.doi.org/10.1086/164633}{{\em The Astrophysical Journal}
  {\bfseries 309} (Oct, 1986) 653}.
  \url{https://ui.adsabs.harvard.edu/abs/1986ApJ...309..653C}.

\bibitem{Hartle:1968si}
J.~B. Hartle and K.~S. Thorne, ``{Slowly Rotating Relativistic Stars. II.
  Models for Neutron Stars and Supermassive Stars},''
  \href{http://dx.doi.org/10.1086/149707}{{\em Astrophys. J.} {\bfseries 153}
  (1968) 807}.

\bibitem{Caldwell:1997ii}
R.~R. Caldwell, R.~Dave, and P.~J. Steinhardt, ``{Cosmological imprint of an
  energy component with general equation of state},''
  \href{http://dx.doi.org/10.1103/PhysRevLett.80.1582}{{\em Phys. Rev. Lett.}
  {\bfseries 80} (1998) 1582--1585},
  \href{http://arxiv.org/abs/astro-ph/9708069}{{\ttfamily
  arXiv:astro-ph/9708069}}.

\bibitem{Kiselev:2002dx}
V.~V. Kiselev, ``{Quintessence and black holes},''
  \href{http://dx.doi.org/10.1088/0264-9381/20/6/310}{{\em Class. Quant. Grav.}
  {\bfseries 20} (2003) 1187--1198},
  \href{http://arxiv.org/abs/gr-qc/0210040}{{\ttfamily arXiv:gr-qc/0210040}}.

\bibitem{PhysRevLett.113.201801}
{\bfseries BaBar Collaboration} Collaboration, J.~P. Lees {\em et~al.},
  ``Search for a dark photon in ${e}^{+}{e}^{\ensuremath{-}}$ collisions at
  babar,'' \href{http://dx.doi.org/10.1103/PhysRevLett.113.201801}{{\em Phys.
  Rev. Lett.} {\bfseries 113} (Nov, 2014) 201801}.
  \url{https://link.aps.org/doi/10.1103/PhysRevLett.113.201801}.

\bibitem{PhysRevLett.120.061801}
{\bfseries LHCb Collaboration} Collaboration, R.~Aaij {\em et~al.}, ``Search
  for dark photons produced in 13 tev $pp$ collisions,''
  \href{http://dx.doi.org/10.1103/PhysRevLett.120.061801}{{\em Phys. Rev.
  Lett.} {\bfseries 120} (Feb, 2018) 061801}.
  \url{https://link.aps.org/doi/10.1103/PhysRevLett.120.061801}.

\bibitem{Ilten:2018crw}
P.~Ilten, Y.~Soreq, M.~Williams, and W.~Xue, ``{Serendipity in dark photon
  searches},'' \href{http://dx.doi.org/10.1007/JHEP06(2018)004}{{\em JHEP}
  {\bfseries 06} (2018) 004}, \href{http://arxiv.org/abs/1801.04847}{{\ttfamily
  arXiv:1801.04847 [hep-ph]}}.

\bibitem{Bjorken:2009mm}
J.~D. Bjorken, R.~Essig, P.~Schuster, and N.~Toro, ``{New Fixed-Target
  Experiments to Search for Dark Gauge Forces},''
  \href{http://dx.doi.org/10.1103/PhysRevD.80.075018}{{\em Phys. Rev. D}
  {\bfseries 80} (2009) 075018},
  \href{http://arxiv.org/abs/0906.0580}{{\ttfamily arXiv:0906.0580 [hep-ph]}}.

\bibitem{Andreas:2012mt}
S.~Andreas, C.~Niebuhr, and A.~Ringwald, ``{New Limits on Hidden Photons from
  Past Electron Beam Dumps},''
  \href{http://dx.doi.org/10.1103/PhysRevD.86.095019}{{\em Phys. Rev. D}
  {\bfseries 86} (2012) 095019},
  \href{http://arxiv.org/abs/1209.6083}{{\ttfamily arXiv:1209.6083 [hep-ph]}}.

\bibitem{Blumlein:2013cua}
J.~Bl\"umlein and J.~Brunner, ``{New Exclusion Limits on Dark Gauge Forces from
  Proton Bremsstrahlung in Beam-Dump Data},''
  \href{http://dx.doi.org/10.1016/j.physletb.2014.02.029}{{\em Phys. Lett. B}
  {\bfseries 731} (2014) 320--326},
  \href{http://arxiv.org/abs/1311.3870}{{\ttfamily arXiv:1311.3870 [hep-ph]}}.

\bibitem{Escudero:2018fwn}
M.~Escudero, S.~J. Witte, and N.~Rius, ``{The dispirited case of gauged
  U(1)$_{B-L}$ dark matter},''
  \href{http://dx.doi.org/10.1007/JHEP08(2018)190}{{\em JHEP} {\bfseries 08}
  (2018) 190}, \href{http://arxiv.org/abs/1806.02823}{{\ttfamily
  arXiv:1806.02823 [hep-ph]}}.

\bibitem{Ghosh:1995dn}
S.~K. Ghosh, S.~C. Phatak, and P.~K. Sahu, ``{Strangeness production in neutron
  stars},'' \href{http://dx.doi.org/10.1016/0375-9474(95)00389-4}{{\em Nucl.
  Phys. A} {\bfseries 596} (1996) 670--683},
  \href{http://arxiv.org/abs/nucl-th/9509008}{{\ttfamily
  arXiv:nucl-th/9509008}}.

\bibitem{Jamil:2014rsa}
M.~Jamil, S.~Hussain, and B.~Majeed, ``{Dynamics of Particles Around a
  Schwarzschild-like Black Hole in the Presence of Quintessence and Magnetic
  Field},'' \href{http://dx.doi.org/10.1140/epjc/s10052-014-3230-7}{{\em Eur.
  Phys. J. C} {\bfseries 75} no.~1, (2015) 24},
  \href{http://arxiv.org/abs/1404.7123}{{\ttfamily arXiv:1404.7123 [gr-qc]}}.

\bibitem{MosqueraCuesta:2017iln}
H.~J. Mosquera~Cuesta, G.~Lambiase, and J.~P. Pereira, ``{Probing nonlinear
  electrodynamics in slowly rotating spacetimes through neutrino
  astrophysics},'' \href{http://dx.doi.org/10.1103/PhysRevD.95.025011}{{\em
  Phys. Rev. D} {\bfseries 95} no.~2, (2017) 025011},
  \href{http://arxiv.org/abs/1701.00431}{{\ttfamily arXiv:1701.00431 [gr-qc]}}.

\end{thebibliography}\endgroup
\end{document}